# Chondrule trace element geochemistry at the mineral scale

Emmanuel Jacquet[1], Olivier Alard[2], Matthieu Gounelle[1,3]

[1]Laboratoire de Minéralogie et Cosmochimie du Muséum, CNRS & Muséum National d'Histoire Naturelle, UMR 7202, 57 rue Cuvier, 75005 Paris, France.
[2]Géosciences Montpellier, UMR 5243, Université de Montpellier II, Place E. Bataillon, 34095 Montpellier cedex 5, France.
[3]Institut universitaire de France, Maisons des Universités, 103 boulevard Saint-Michel, 75005 Paris, France.

E-mail: ejacquet@mnhn.fr

*Abstract*

We report trace element analyses from mineral phases in chondrules from carbonaceous chondrites (Vigarano, Renazzo and Acfer 187), carried out by laser ablation inductively coupled plasma mass spectrometry. Results are similar in all three meteorites. Mesostasis has Rare Earth Element (REE) concentrations of 10-20 x CI. Low-Ca pyroxene has light REE (LREE) concentrations near 0.1 x CI and heavy REE (HREE) near 1 x CI respectively. Olivine has HREE concentrations at 0.1-1 x CI and LREE around $10^{-2}$ x CI. The coarsest olivine crystals tend to have the most fractionated REE patterns, indicative of equilibrium partitioning. Low-Ca pyroxene in the most pyroxene-rich chondrules tends to have the lowest REE concentrations. Type I chondrules seem to have undergone a significant degree of batch crystallization (as opposed to fractional crystallization), which requires cooling rates slower than 1-100 K/h. This would fill the gap between igneous CAIs and type II chondrules. The anticorrelation between REE abundances and pyroxene mode may be understood as due to dilution by addition of silica to the chondrule melt, as in the gas-melt interaction scenario of Libourel et al. (2006). The rapid cooling rate (of the order of 1000 K/h) which seems recorded by low-Ca pyroxene, contrasted with the more diverse record of olivine, may point to a nonlinear cooling history or suggest that formation of pyroxene-rich chondrule margins was an event distinct from the crystallization of the interior.

## *1. Introduction*

Chondrules are the major components of primitive meteorites and likely formed by brief melting of precursor material followed by rapid cooling and crystallization. However ubiquitous they are, their formation, presumably an important astrophysical process in the early inner solar system, is still largely shrouded in mystery. Candidate heating/forming mechanisms, such as gas-liquid condensation (e.g. Blander et al. 2004; Varela et al. 2006), the shock-wave models (Desch et al. 2005), collision with molten planetesimals (e.g. Asphaug et al. 2011), etc. are actively being debated (Connolly and Desch 2004; Ciesla 2005). Distinguishing between these scenarii requires clear cosmochemical constraints on the nature of the precursors, the thermal history of chondrules, the behavior of volatile elements, the genetic relationship between chondrules and matrix (Hezel and Palme 2010; Hezel and Palme 2008; Zanda et al. 2006) and the diversity of chondrule reservoirs sampled by chondrites.

It is traditionally assumed that chondrule precursors were millimeter-sized "dustballs", but a significant proportion of coarse components are needed to account for the variability of chondrule bulk composition (Hezel and Palme 2007). The precursors could be either (i) nebular condensates, (ii) planetesimal debris (iii) a previous generation of chondrules or a combination thereof. As to (i), rare relict Calcium-Aluminum-rich Inclusions (CAIs) have been observed in chondrules (Krot et al. 2009) and some Rare Earth Element (REE) patterns of bulk chondrules seem to be inherited from such precursors (e.g. Pack et al. 2004). Regarding (ii), olivine-rich lithic inclusions with 120° triple junctions, known as Granoblastic Olivine Aggregates (GOA), could be precursors of porphyritic olivine (PO) chondrules (e.g. Libourel and Krot 2007; Whattam and Hewins 2009), and were interpreted by Libourel and Krot (2007) to be mantle debris of differentiated planetesimals. Besides, rare achondritic clasts have been described in chondrites (Mittlefehldt et al. 1998). Finally, (iii) is supported by the frequent occurrence of relict olivine crystals in chondrules that presumably formed under different oxygen fugacity conditions than their host (Jones 1996), the existence of compound chondrules and that of coarse-grained (igneous) rims around many chondrules (Rubin and Krot 1996). Ruzicka et al. (2008) argued that multiple heating events allowed type I chondrules (i.e. chondrules with less than 10 wt% FeO in olivine) or chondrule fragments to be remelted under oxidizing conditions, leading to type II chondrules (with more than 10 wt% FeO in olivine), and vice-versa. However, Hezel and Palme (2007) pointed out that this recycling must be limited in extent to account for chondrule variability.

The thermal history of chondrules has been constrained by experiments (Hewins et al. 2005). Preservation of sulfides as well as olivine nuclei limits heating times near olivine/silicate liquidus temperatures (1400-1750 °C) to a few minutes at most (Hewins et al. 2005). While chondrule textures are consistent with a wide array of cooling rates (4-3000 K/h as reported by Hewins et al. 2005), olivine zoning in type II chondrules favor the 100-1000 K/h range (Hewins et al. 2005).

Experiments and theory suggest that chondrule-forming events should be accompanied by significant evaporation of volatile elements such as sulfur or alkalis, and indeed, the depletion of volatile elements in bulk chondrules correlates with grain size (Hewins et al. 1997; Gordon 2009). However, no significant isotope fractionation has been measured for those elements (Davis et al. 2005) and Alexander et al. (2008) found that sodium content was relatively constant between the onset and the end of olivine crystallization. Cuzzi and Alexander (2006) suggested that this could be due to equilibrium with gas evaporated from neighboring chondrules, implying a high number density ($> 10$ $m^{-3}$) of the latter during chondrule-forming events. Evidence for open-system behavior for type I chondrules was presented by Tissandier et al. (2002) and Libourel et al. (2006), who ascribed the formation of pyroxene near chondrule margins to gas-melt interaction, which would add silica to the melt (see also Hezel et al. (2003) for silica-rich components in CH chondrites).

Trace element geochemistry is potentially a powerful tool to shed new light on these issues. Indeed trace elements display a wide diversity of geochemical behaviors, and their abundance is sensitive to formation conditions at the order-of-magnitude level. Trace element concentrations, in particular for Rare Earth Elements (REE), of bulk chondrules have been determined in a number of studies (e.g. Misawa and Nakamura 1988; Pack et al. 2004; Pack et al. 2007; Inoue et al. 2009; Gordon 2009) with REE generally found to be essentially unfractionated ($0.7 \leq (Ce/Yb)_N \leq 1.2$ for Misawa and Nakamura (1988); the N subscript denotes normalization to CI1 chondrites, values after Lodders 2003).

To study chondrules as igneous systems, it is important to analyze trace elements for individual mineral phases in chondrules. Secondary Ion Mass Spectrometry (SIMS) data for olivine were reported by Alexander (1994) and Ruzicka et al. (2008), who studied L and LL chondrites, respectively, with REE found to be weakly fractionated ($0.2 \leq (Ce/Yb)_N \leq 1.3$, excluding one olivine at 2.3, for Alexander 1994), and interpreted to indicate cooling rates of order 1000 K/h. Alexander (1994) and Jones and Laynes (1997) found low-Ca pyroxene to be weakly fractionated in L chondrites ($0.4 \leq (Ce/Yb)_N \leq 1.3$) and Semarkona ($0.06 \leq (Ce/Yb)_N \leq 0.16$), respectively.

In this paper, we report and discuss *in situ* Laser Ablation Inductively Coupled Plasma-Mass Spectrometry (LA-ICP-MS) analyses of mineral phases (olivine, pyroxene, mesostasis) of 36 chondrules and isolated grains in three carbonaceous chondrites, from the CV and CR groups. To our knowledge, this is the first time that this technique, which allows rapid measurements compared to SIMS, is applied to individual chondrule silicates. Also, while the above trace element studies focused on ordinary chondrites, our work contributes data for a qualitatively different chondrite "super-clan" (Kallemeyn et al. 1996). Carbonaceous chondrites contain a distinctive chondrule population, dominated by type I, while type II chondrules are common in ordinary chondrites. The CV and CR groups are most suitable for our analyses since they both contain relatively large, unaltered chondrules. These two groups are at the same time very distinct in the whole spectrum of carbonaceous chondrites: CV are the most CAI- and $^{16}$O-rich chondrites, while CR chondrites, poor in CAIs but endowed with characteristic metal-mantled chondrules and the most primitive insoluble organic matter (Alexander 2005), are the most $^{16}$O-poor carbonaceous chondrites, excluding (aqueously altered) CMs and CIs. Thus, our analyses allow a first glimpse into the diversity of carbonaceous chondrites as regards trace element microdistribution in chondrules and the constraints it provides on chondrule-forming processes.

## *2. Samples and analytical procedures*

Three polished sections from the meteorite collection of the Muséum National d'Histoire Naturelle de Paris (MNHN) were used in this study: section 2006-14 of Vigarano (CV3), section 719 of Renazzo (CR2), both observed falls, and the section of Acfer 187 (CR2), a find paired with El Djouf 001. The sections were examined in optical and scanning electron microscopy (SEM—here a

JEOL JSM-840A instrument). X-ray maps allowed apparent mineral modes of chondrules to be calculated using the JMicrovision software (www.jmicrovision.com). Minor and major element concentrations of documented chondrules were obtained with a Cameca SX-100 electron microprobe (EMP) at the Centre de Microanalyse de Paris VI (CAMPARIS), using well-characterized mineral standards. Analysis spots were 5 µm in diameter and the beam current and accelerating voltage were 10 nA and 15 kV, respectively. Counting time was 10 s for Mg, Ca, Al, S, Cr, Mn, Ti, Ni, P, Co and Fe and 5 s for Si, Na and K.

Trace element analyses of selected chondrules were performed by LA-ICP-MS at the University of Montpellier II. The laser ablation system was a GeoLas Q$^+$ platform with an Excimer CompEx 102 laser and was coupled to a ThermoFinnigan Element XR mass spectrometer. The ICP-MS was operated at 1350 W and tuned daily to produce maximum sensitivity for the medium and high masses, while keeping the oxide production rate low ($^{248}$ThO/$^{232}$Th ≤ 1%). Ablations were performed in pure He-atmosphere (0.65 ± 0.05 l•min$^{-1}$) mixed before entering the torch with a flow of Ar (≈ 1.00 ± 0.05 l•min$^{-1}$). Laser ablation condition were: fluences ca. 12J/cm² with pulse frequencies between 5 and 10 Hz were used and spot sizes of 25-100 µm, 50 µm being a typical value. With such energy fluences, depth speed for silicates is about 1 µm•s$^{-1}$. Each analysis consisted of 4 min on background analyses (laser off) and 1 min of ablation (laser on). Long background times and short analysis times were chosen in order to have a statistically meaningful analysis of a low background noise. In order to further minimize background noise, analyses in each chondrule were performed in the order of increasing incompatible element abundances (olivine < low-Ca pyroxene < high-Ca pyroxene < mesostasis). Data reduction was carried out using the GLITTER software (Griffin et al. 2008). Internal standard was Si (Ca for augite), known from EMP analyses. The NIST 612 glass (Pearce et al. 1997) was used as an external standard. This double standardization allows correction for variations in ablation yield and instrumental drift (Longerich et al. 1996). The time-resolved nature of these analyses and the GLITTER software allow one to detect (and avoid in the signal selection) transitions to phases other than that of interest during ablation, as well as inclusions, down to a scale corresponding to one mass scan, about a micrometer.. The absence of phase overlap in the transverse direction was also controlled by direct SEM examination, previous back-scattered electron images also enabling one to check the cleanness of the analyzed area. If a spatial precision of one micrometer is conservatively assigned to both signal filtering and SEM observation, the contamination by one single undetected inclusion should not exceed 10$^{-4}$ for a 50 µm-spot and 30 s-signal, but of course the contribution could be higher if the impurities occur as finely dispersed (submicrometer-sized) bodies throughout the analyzed phase. LA-ICP-MS analyses that did not substantially agree with EMP results (in particular for Al, Ca) for the same crystals were discarded. Some trace element analyses of mesostasis (which in all cases included both glass and embedded crystallites) were performed by SIMS at the University of Montpellier II. For each chondrule, a geometric (rather than arithmetic) averaging was used to calculate mean concentrations in olivine and pyroxene in order to further reduce the impact of possible contamination by incompatible element-rich phases. In the following text and in the figures, and unless otherwise noted, the data reported will be *chondrule means*, that is, for each phase, the average of the different analyses performed on that phase in a given chondrule.

## *3. Results*

### 3.1 Petrography and mineralogy

In this section we will briefly describe the petrography of analyzed chondrules, some images of which are provided in Fig. 1. In this article we shall use the word "chondrule" in a wide sense encompassing all silicate-rich objects with petrographic evidence of melting regardless of their overall shape and possible fragmentary nature. Chondrule names begin with a "V", "R" or "A" depending on whether they belong to Vigarano, Renazzo and Acfer 187, respectively. Almost all analyzed chondrules were of type I, with the exception of one type II chondrule per meteorite.

Most type I chondrules (mean Fa content: 3 mol%, diameter between 200 µm and 6 mm) analyzed in Vigarano belong to the microtextural continuum between GOA and porphyritic chondrules (Whattam and Hewins 2009), with olivine, olivine-pyroxene and pyroxene-rich varieties (noted PO, POP and PP, respectively), the latter including the 5 mm chondrule V3 depicted in Fig. 1f, g. LA-ICP-MS analyses were also obtained for three barred olivine (BO) chondrules found in this section. In addition, we analyzed isolated olivine grains (henceforth IOG), including two refractory forsterites (V49 (Fig. 1a) and V54, with Fo content up to 99.8 mol%), and a relatively ferroan olivine (V51, with $Fo_{88-95}$). Olivine aggregates (mostly around $Fo_{99}$) show a microtextural gradation between 120° triple junctions and mere close packing, and the occurrence of triple junctions extends to relatively fine-grained aggregates (< 20 µm crystals, which were unsuitable for LA-ICP-MS analyses). Most GOAs are irregular in shape but some are spheroidal (e.g. chondrule V4, with a granoblastic crust, cf Fig. 1c, d). Low-Ca pyroxene, poikilitically enclosing rounded olivine crystals and often overgrown by augite, generally occurs at the chondrule margins (which is in particular true of many isolated olivine grains), but enstatite blades may extend to the interior. Only rarely is low-Ca pyroxene seen in non-PP chondrule center with no connection to the periphery in the plane of the section. Fe-Ni metal is the dominant opaque mineral but sulfides and magnetite also occur. Chondrule mesostases are only rarely fully glassy and generally exhibit quench augite or plagioclase crystallites, or may even be holocrystalline. 1-20 µm melt inclusions are common in olivine (see also Tschermak 1885). Although Vigarano belongs to the reduced CV subgroup, some minor alteration is visible, with some mesostasis being replaced by phyllosilicates and silicates are often cut by iron-rich veinlets. The only type II chondrule (V48) found in the section was also analyzed.

Notwithstanding the frequent discontinuous metal layers typical of CR chondrites, type I chondrules of Renazzo (mean Fa content: 2 mol%) are broadly qualitatively similar to those of Vigarano. Olivine crystals in PO chondrules seem more frequently euhedral than in Vigarano. No GOA was identified in the section (consistent with their relative paucity mentioned by Whattam and Hewins 2009), although triple junctions are not uncommon. Several type II chondrules were found and one was analyzed (R2); one glass-rich chondrule was also analyzed (R16; cf Fig. 1k). Alteration is more extensive than in Vigarano: In particular type II chondrule mesostases are largely replaced by phyllosilicates and several pyroxene crystals are entirely altered around olivine chadacrysts. Many chondrules are somewhat flattened and roughly exhibit a common orientation in the section (see also Rubin and Swindle 2011).

Chondrules in Acfer 187 are overall larger (0.8-3 mm, n=22) than in Renazzo (0.5-3 mm, n=46), with our section of Acfer 187 containing a dozen chondrules exceeding 1 mm in diameter. Silica pods are common in type I chondrule rims. Al-rich chondrules are fairly abundant. One metal-rimmed granoblastic chondrule (A8; cf Fig. 1m), with incipient disaggregation of triple junctions in low-Ca pyroxene, was found and analyzed. Aqueous alteration seems less intense in Acfer 187 (e.g. the mesostasis of type II chondrule A15 (Fig. 1n) is largely intact) than in Renazzo but oxide-filled veins due to weathering crosscut the section.

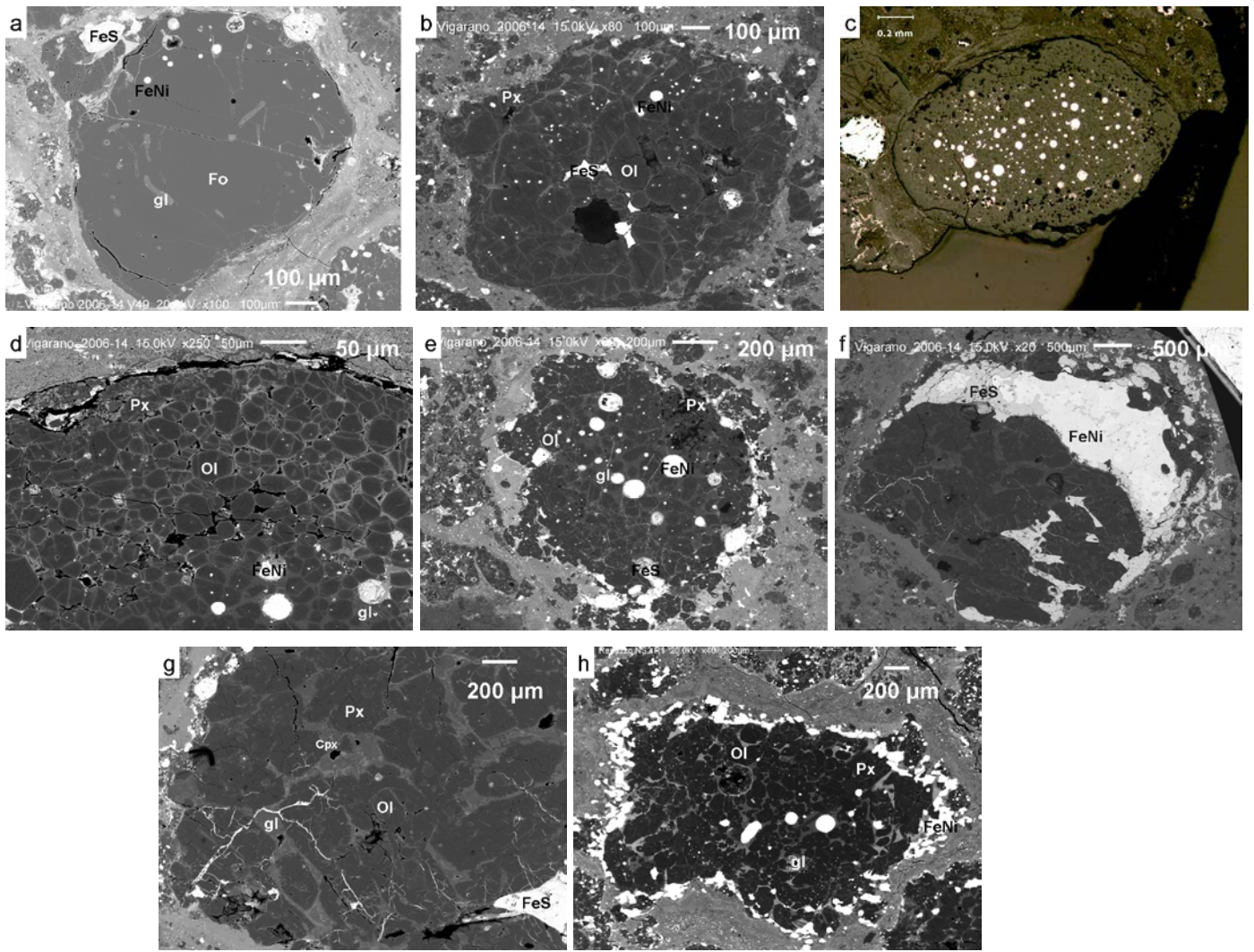

**Figure 1**: Images of chondrules (Back-Scattered Electron images unless otherwise noted).

*Vigarano chondrules.* a) V49 is essentially a single, 0.9 mm refractory forsterite crystal (Fo$_{99.6-99.8}$), with a thin (generally < 20 µm, except in the upper left corner) enstatite shell. Included in the olivine are worm-shaped devitrified mesostasis inclusions (with one spinel-rich in the upper right center) and Fe-Ni metal nodule (some enclosed in the aforementioned mesostasis "worms"). The two biggest nodules are troilite. b) Chondrule V6 is composed of rounded interlocking olivine crystals, with little mesostasis and a thin pyroxene outer layer. Elongate opaque grains are troilite and the rounded ones (mostly intragranular) are iron-nickel metal. c) Reflected-light photomicrograph of chondrule V4. This ovoid (1.2 x 0.8 mm) object exhibits a 200 µm thick granoblastic crust while the interior has an essentially PO texture, with mesostasis and Fe-Ni nodules (both lacking in the periphery). d) Close-up of the granoblastic crust of chondrule V4 (upper part), with transition to the porphyritic interior. A thin (< 30 µm in maximum thickness) enstatite layer overlies the chondrule. e) Chondrule V8 is a PO chondrule, with rounded olivine crystals, a pyroxene-rich rim, iron-nickel nodules and elongate sulfide grains, and a largely devitrified mesostasis. f) V3 is a giant (5 mm diameter) PP chondrule, with olivine only occurring as rounded chadacrysts in the enstatite oikocrysts (typically ~500 µm across). A large sulfide-metal shell is obvious in the upper half, with some Ni-rich zones in its central regions. g) Close-up in chondrule V3. The mesostasis, bounded by augite overgrowths on enstatite, has sparse crystallites, hence its snowy appearance. The cracks are filled with sulfide and do not extend outside the chondrule and are thus probably pre-accretionary.

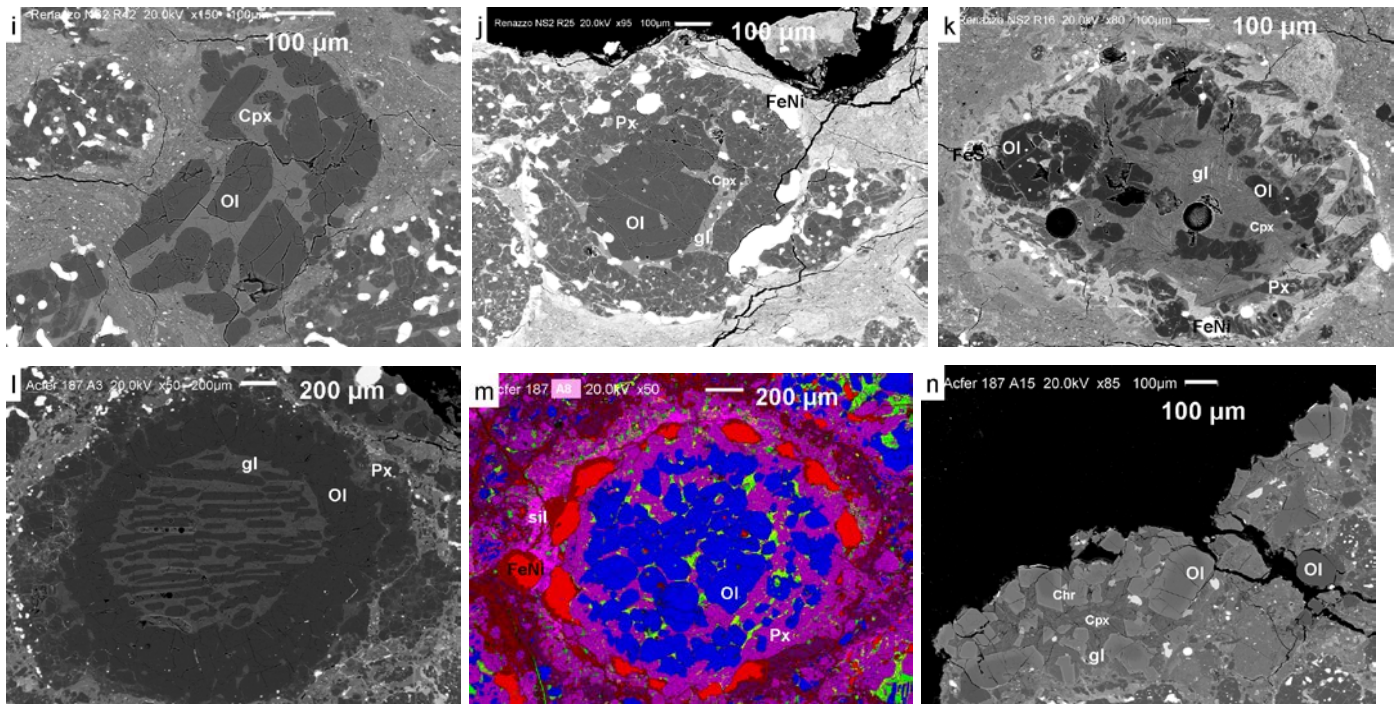

**Figure 1:** *(continued)* Images of chondrules (Back-Scattered Electron images unless otherwise noted).

*Renazzo chondrules.* h) R1 is a PO chondrule of elongate shape, with subhedral olivine crystals, large pyroxene blades concentrated at the margin and finely devitrified mesostasis partly replaced by phyllosilicates near the edge. Fe-Ni metal occurs as nodules in the interior and as a discontinuous outermost shell. A hole due to the polishing process appears near the center. i) R42 is a POP chondrule with olivine phenocrysts set in augite. j) R25 is essentially a single large olivine crystal, 400 µm in diameter, surrounded by a thick (100-200 µm) enstatite shell loaded with Fe-Ni grains and olivine chadacrysts. Sandwiched between the olivine and the enstatite is a <50 µm thick mesostasis-augite layer. k) R16 is a glass-rich ovoid chondrule, with devitrified mesostasis, subhedral low-Ca pyroxene crystals overgrown by augite, and anhedral olivine grains. The outer part of the chondrule is altered, and near the alteration front, the mesostasis is enriched in Na. On the left, an ovoid PO chondrule is entirely enclosed in the chondrule. Opaque nodules are Fe-Ni metal, except the leftmost one (troilite). The two round holes are previous LA-ICP-MS spots.

*Acfer 187 chondrules*. l) A3 is a 2.2 mm-diameter BO chondrule, with a 500 µm-thick outer shell and 30-50 µm-broad olivine bars, set in a devitrified mesostasis (with euhedral augite crystallites). Enstatite makes up a thin (< 100 µm) outer layer, with interspersed Fe-Ni specks. m) Combined X-ray map of chondrule A8. Blue = Mg, pink = Si, red = Fe, green = Ca, yellow = Al. The interior is granoblastic in texture, with some infiltration of enstatite, predominantly near the margin. A discontinuous metal shell is apparent inside the pyroxene-rich rim, and silica occurs beyond it. n) A15 is a type II PO chondrule (interrupted by the section's edge). Relatively large subhedral olivine grains are set in a devitrified mesostasis, which encloses augite, chromite and troilite. A nearby isolated magnesian olivine grain (also marked "Ol") was also analyzed.

Ol = olivine, Px = low-Ca pyroxene, Cpx = Ca-rich pyroxene, gl = mesostasis, FeS = troilite, FeNi = iron-nickel alloy, sil = silica, Chr = chromite.

## 3.2 Chemistry

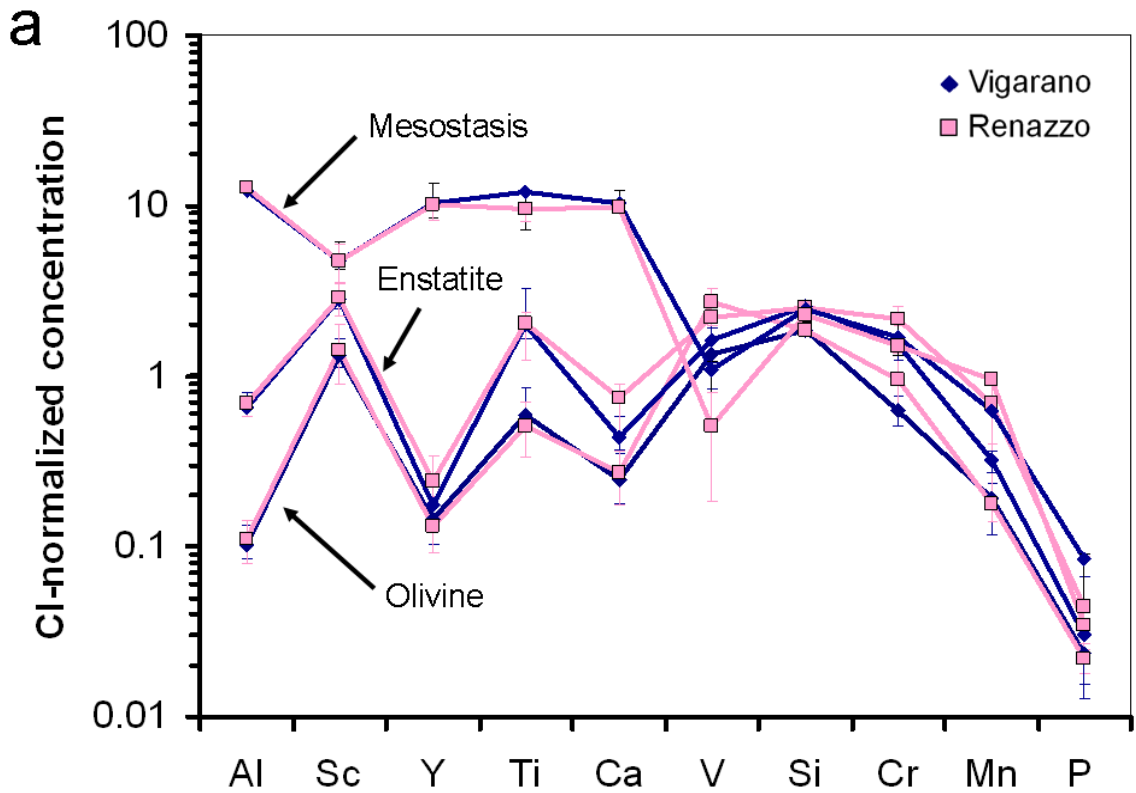

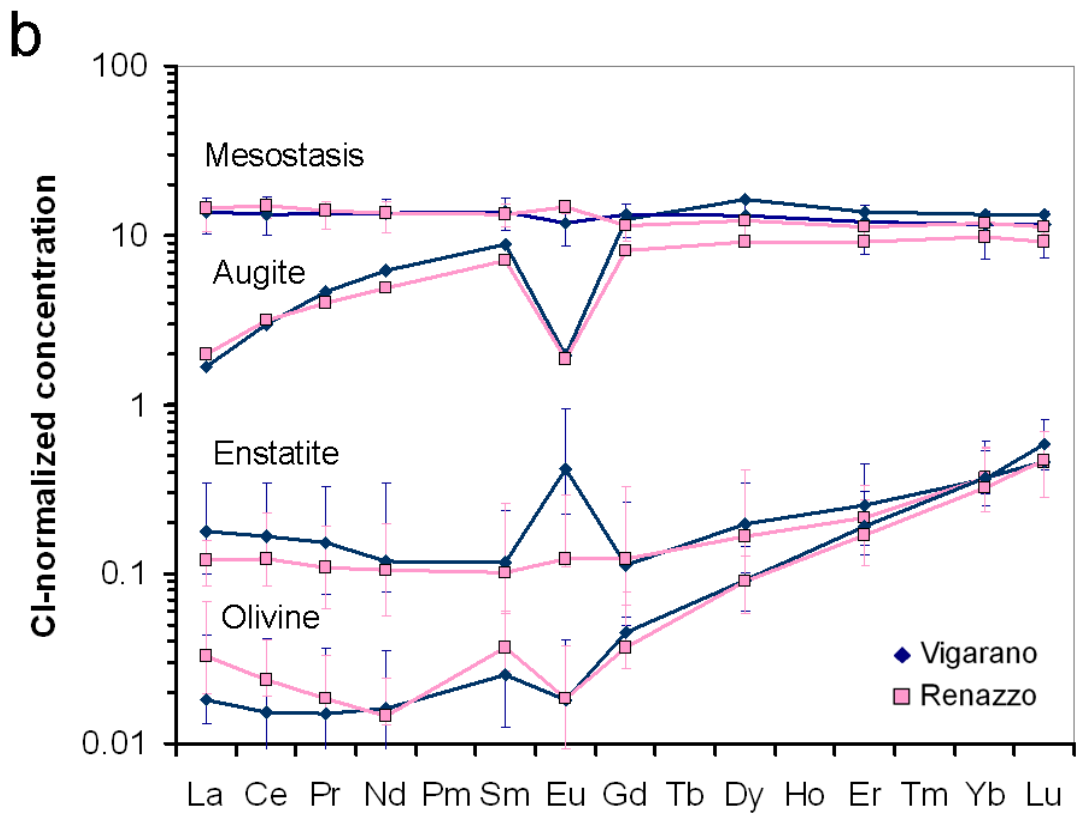

**Figure 2:** (a) CI-normalized average concentrations of lithophile elements arranged according to volatility for olivine, enstatite and mesostasis in type I chondrules in Vigarano and Renazzo (the average being performed on chondrule means). The bars indicate the first and third quartiles of the data set. (b) Same for Rare Earth Elements (including the mineral augite; with no bar due to poor statistics). The number of chondrules averaged for each phase are (with in parentheses the corresponding total number of *analyses*

involved): Olivine: 20 (50) and 9 (23), enstatite: 9 (14) and 5 (7), augite: 2 (2) and 3 (6), mesostasis: 10 (20) and 5 (10), for Vigarano and Renazzo, respectively.

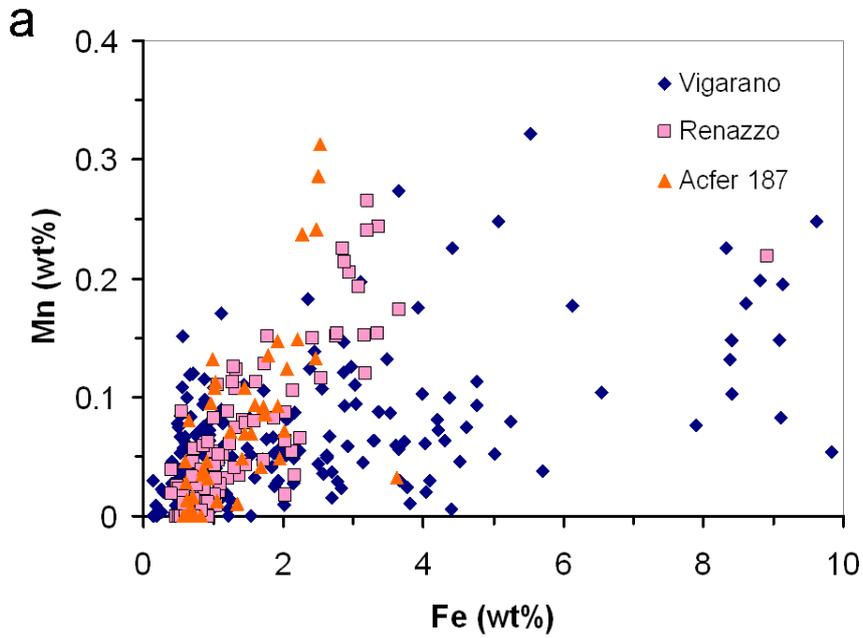

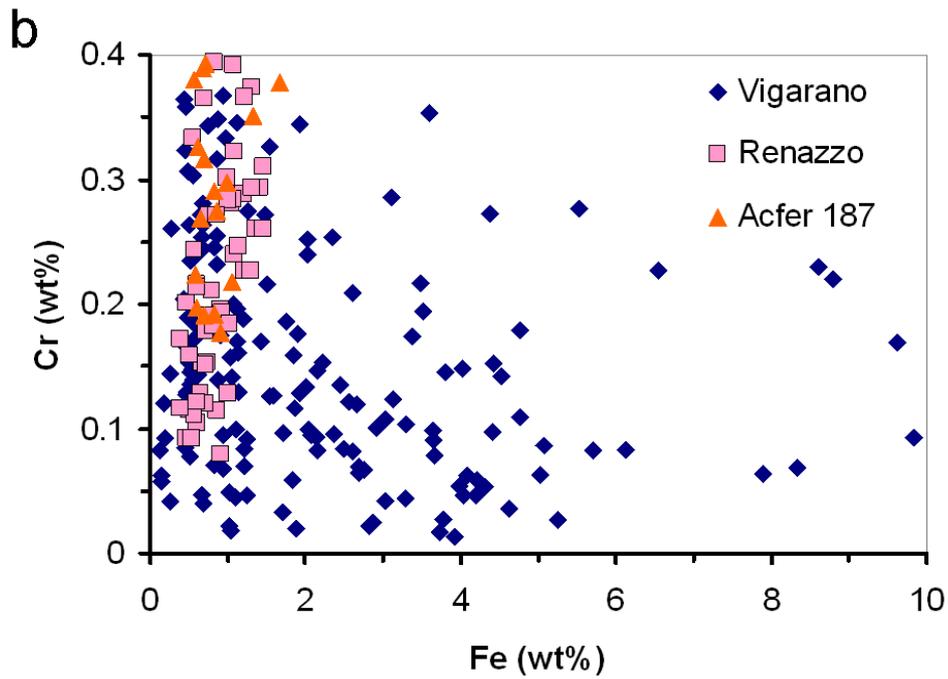

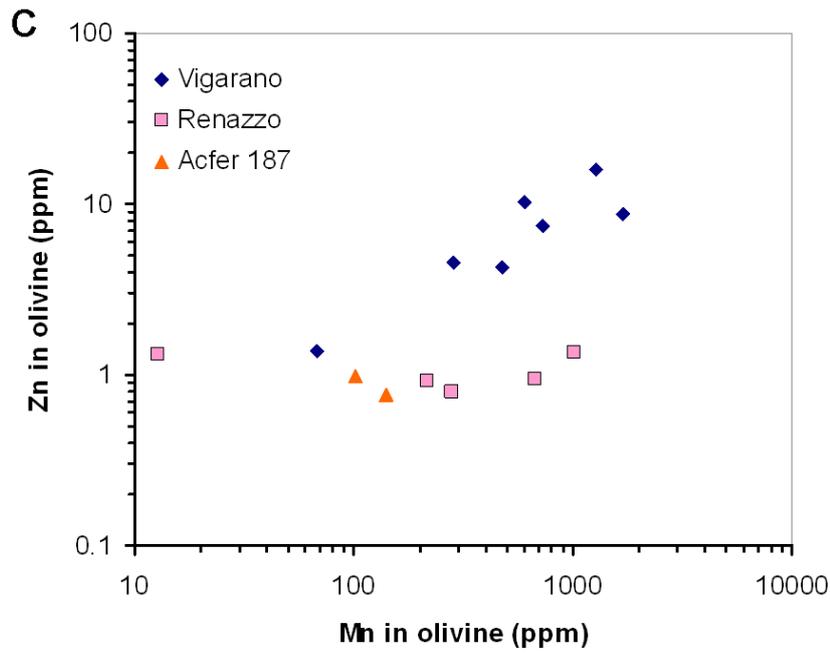
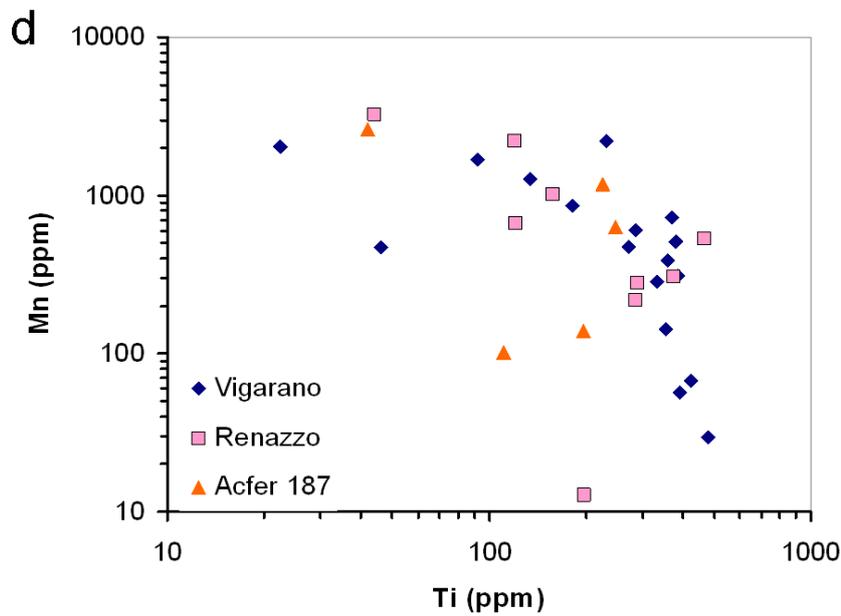

**Figure 3:** Minor and trace element plots in type I chondrule olivine. (a) Mn vs. Fe (b) Cr vs. Fe (c) Zn vs. Mn (d) Mn vs. Ti. Plots (a) and (b) are based on electron microprobe data on individual crystals while plots (c) and (d) are based on LA-ICP-MS data, with each point representing one chondrule mean.

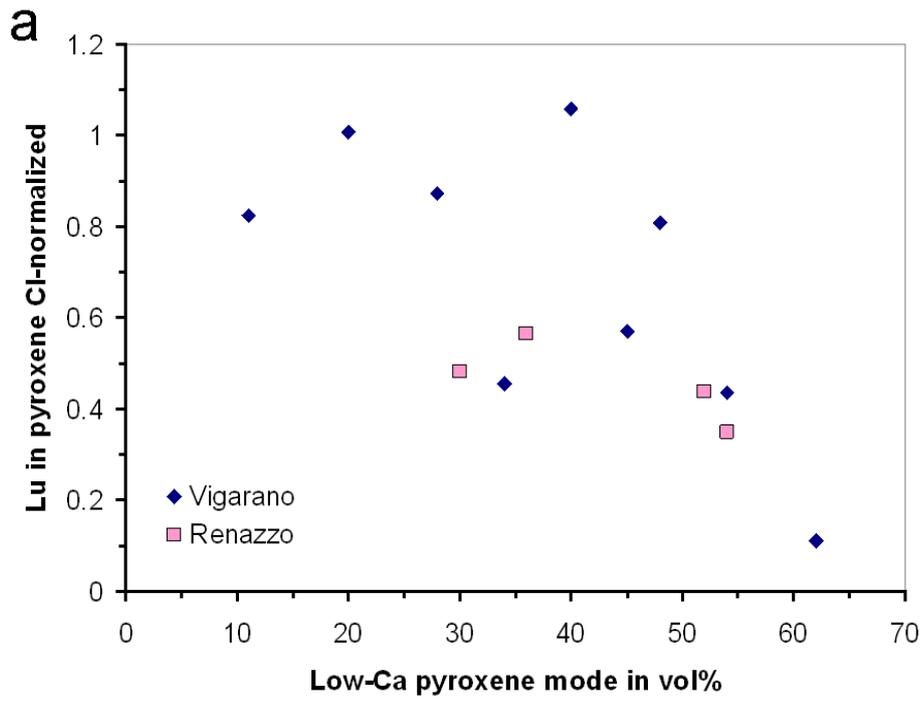

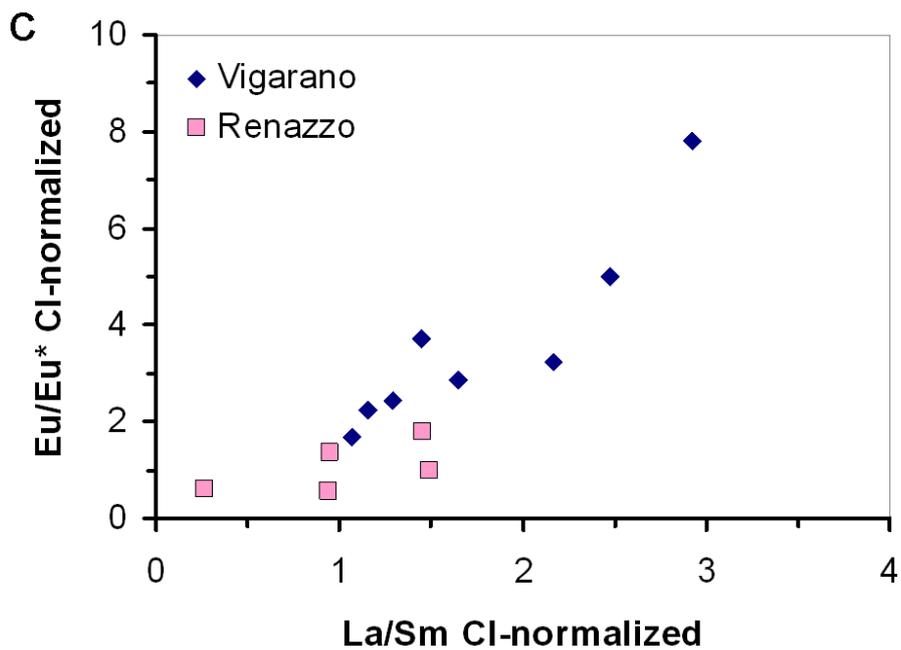

**Figure 4:** Plot of lutetium concentration in low-Ca pyroxene (a) and mesostasis (b) versus low-Ca pyroxene mode in type I chondrules. (c) Plot of Eu/Eu* (the europium anomaly) as a function of La/Sm (CI-normalized) in low-Ca pyroxene.

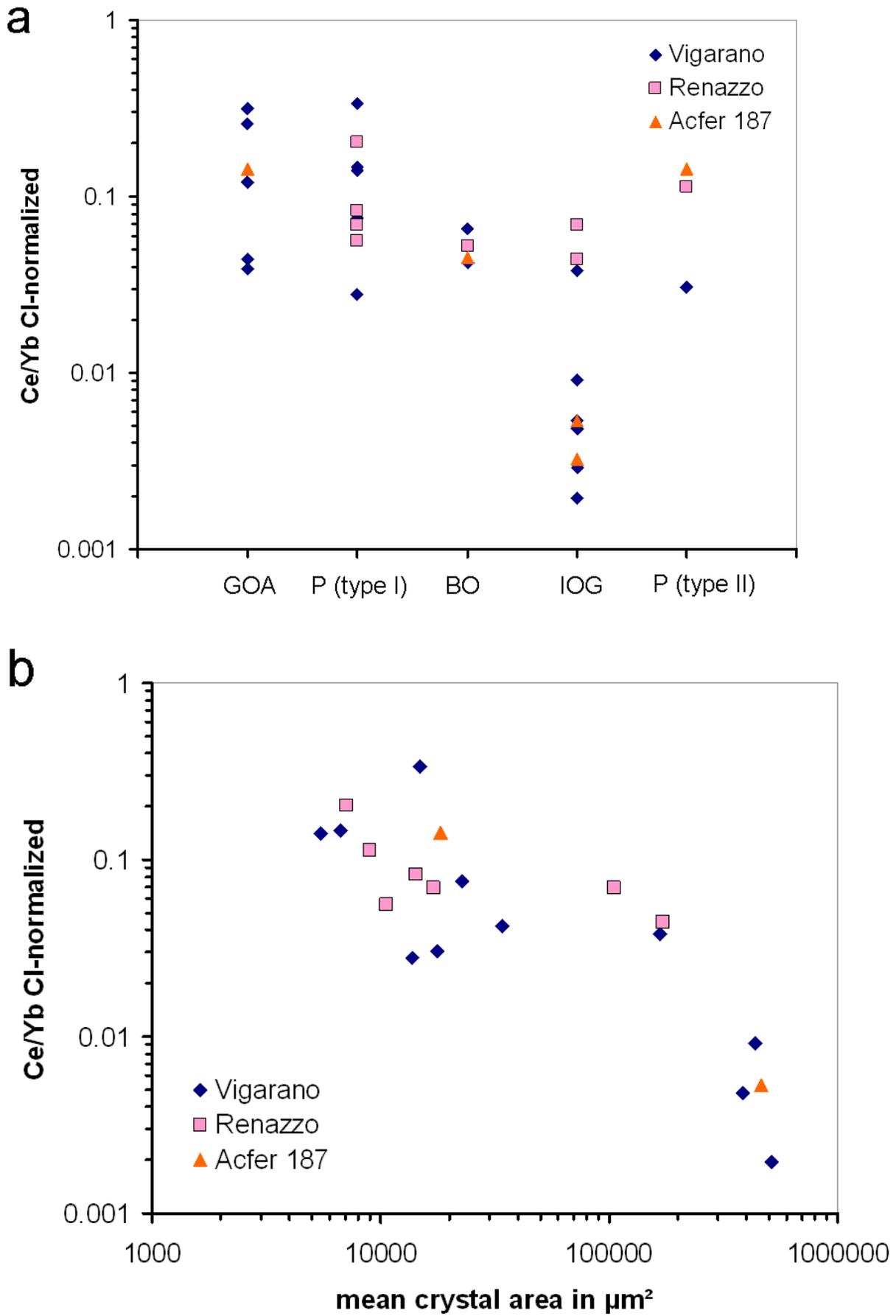

**Figure 5:** (a) CI-normalized Ce/Yb ratio of chondrule olivine as a function of textural type: GOA = granoblastic olivine aggregate, P = porphyritic, BO = barred olivine, IOG = isolated olivine grain. Type II

chondrules analyzed here are PO chondrules. (b) Plot of $(Ce/Yb)_N$ in olivine as a function of mean crystal cross-sectional area for porphyritic chondrules and unbroken IOG. Each data point represents one chondrule, for which the $(Ce/Yb)_N$ of the mean olivine composition is used, and the mean area (as measured on BSE images) is the average of that of the crystals used to calculate the mean. This averaging removes some scatter (e.g. due to nonequatorial sectioning of olivine crystals or analytical uncertainties of the LA-ICP-MS).

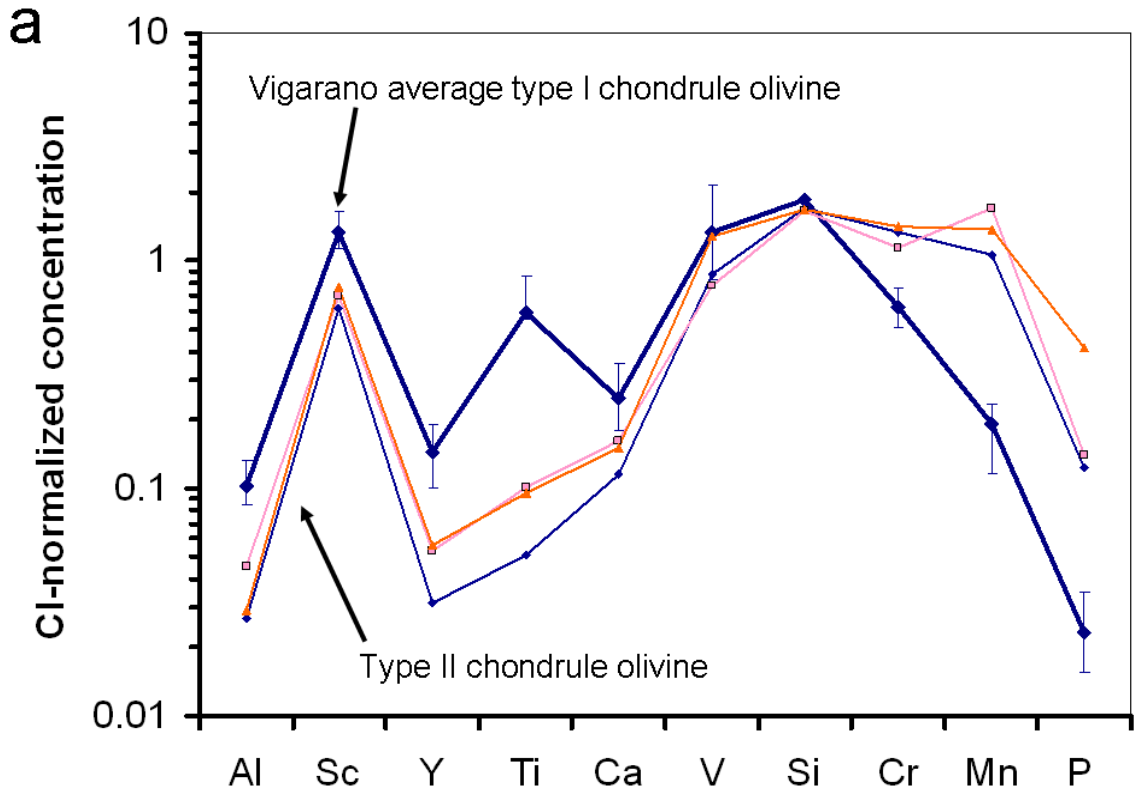

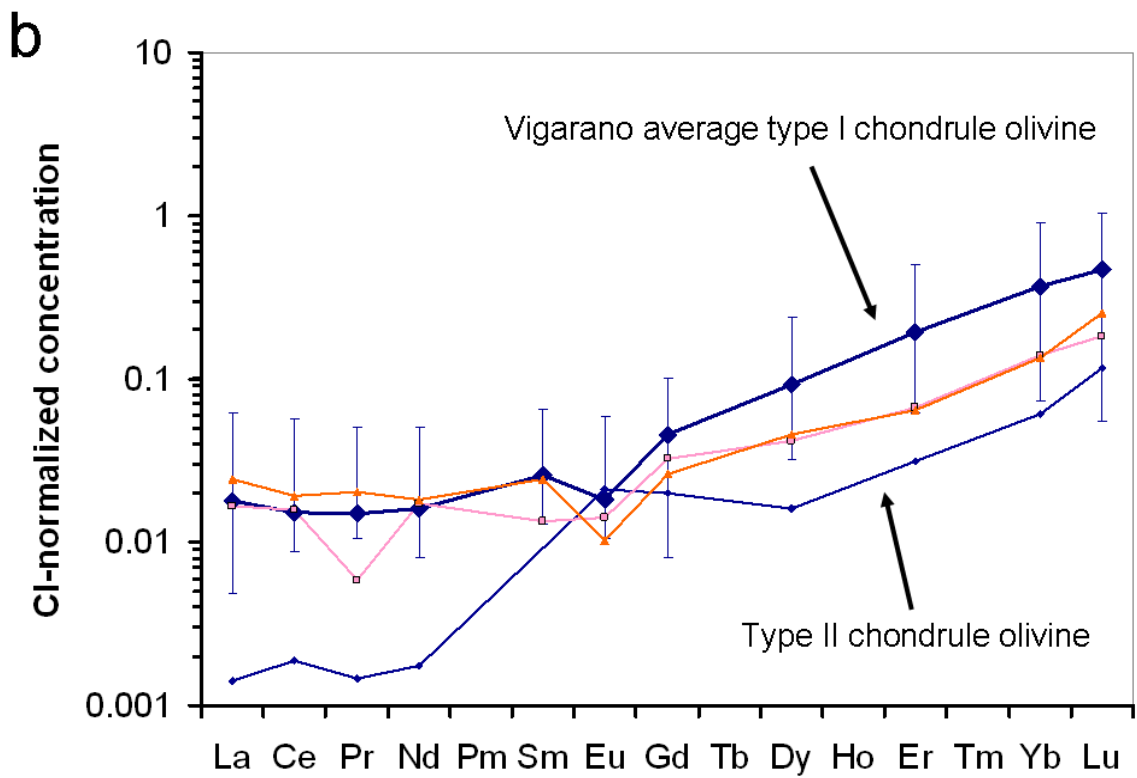

**Figure 6:** Trace element patterns for type II chondrule olivine in Vigarano, Renazzo and Acfer 187 (one in each meteorite) as compared to type I olivine in Vigarano (that of Renazzo, very similar to the former, being omitted for clarity) for the same sets of elements presented in Fig. 2.

Trace element data for each chondrite and each phase, averaged over all type I chondrule (the vast majority of chondrules in CV and CR chondrites) are reported in Table 1 and displayed in Fig. 2 for two sets of elements, including Rare Earth Elements (individual chondrule data are given in the Electronic Annex). In general, trace and minor element concentrations in olivine, pyroxene and mesostasis are similar in the three different meteorites studied, in terms of the ranges spanned or interelement correlations. As regards olivine — the phase with the most extensive data set —, that of type I chondrules of Vigarano seems systematically enriched in Zn and Pb compared to Renazzo (4 and 0.3 ppm vs 1 and 0.01 ppm on average); Mn and Cr correlate with Fe more strongly in CRs than in Vigarano (see also Brearley and Jones 1998) but Mn in Vigarano is positively correlated with Zn and negatively correlated with Ti (see Fig. 3). The scatter in the Cr vs. Fe plot (Fig. 3b) in Vigarano could be ascribed to chromite precipitation due to mild metamorphism (Grossman and Brearley 2005). Otherwise, the data appear to be very similar—at least in terms of the correlations that they define—from meteorite to meteorite, and we shall thus mostly discuss the results collectively henceforth.

We now focus on REE abundances of the different mineral phases in chondrules (see Fig. 2b). We will denote by LREE, MREE and HREE the light, middle and heavy REE, respectively. We will draw comparisons with literature data mentioned in the introduction but one should keep in mind that the latter bear on *ordinary* chondrites while the analyses reported here involve CR and CV chondrites.

Mesostases have nearly flat, REE patterns showing a steady weak enrichment from the HREE to the LREE ($1 \leq (Ce/Yb)_N \leq 1.7$). La concentration ranges from 6 to 22 x CI. Weak positive and negative europium anomalies occur, with $Eu_N/Eu^*_N$ ranging between 0.7 and 1.6 ($Eu^*_N = (Sm_N + Gd_N)/2$). These data agree well with SIMS measurements performed on the mesostasis of ordinary chondrite chondrules (Alexander 1994; Jones and Laynes 1997; Ruzicka et al. 2008).

Plagioclase crystals in holocrystalline mesostasis were analyzed in two chondrules V52 and R12 ($An_{78}$ and $An_{100}$, respectively). In each case, the REE abundance patterns show LREE enrichments, with a positive Eu anomaly (($Ce/Yb)_N$ = 3.4 and 4.5, Eu/Eu* = 5 and 19 and La concentration 3.7 and 1.0 x CI, respectively). This is similar to plagioclase analyses carried out by Jones et al. (2001) in an Al-rich chondrule from Mokoia (CV3).

For augite Lu varies between 6-20 x CI and La is ca. 0.8-2 x CI (except chondrule R42, at 12 x CI). Augites show HREE enrichments ($0.2 < (Ce/Yb)_N < 0.9$) and negative europium anomalies with Eu/Eu* between 0.1 and 0.5 and a flattening for the HREE. Here again the agreement with the literature data (Alexander 1994; Jones and Laynes 1997) is excellent.

Low-Ca pyroxene shows much lower REE content than high Ca Pyroxenes ($0.1 < Lu_N \leq 1.9$; $0.05 < La_N \leq 0.5$). The two PP chondrules show the lowest REE concentrations. CI-normalised REE abundances for low-Ca pyroxenes steadily decrease from HREE to MREE ($1.8 < (Lu/Sm)_N \leq 25.3$) and then flatten for the LREE ($0.8 < (La/Sm)_N < 2.9$; $0.1 \leq (Ce/Yb)_N \leq 0.7$). Positive Eu anomalies are common, especially in Vigarano (Eu/Eu* = 0.6-8), and correlate with the La/Sm ratio (Fig. 4c). Two pigeonites display REE patterns intermediate between those of low- and high-Ca pyroxenes. These data agree well with SIMS data reported by Alexander (1994) but analyses by Jones and Laynes (1997) yielded systematically lower REE abundances, by one order of magnitude for LREE. However, we note that REE content of enstatite (as well as of mesostasis) seems to be anticorrelated with pyroxene mode (Fig. 4a,b). Inasmuch as the chondrules analyzed by Jones and Laynes (1997) were pyroxene-rich, the discrepancy between their data and those of Alexander (1994) could be ascribed to a petrographic bias.

Olivines have the lowest REE contents, with Lu at 0.1-1.3 x CI while La occurs typically at a few ppb level (ranging from 0.0008 to 0.15 x CI). Thus LREE are strongly depleted relative to HREE with $(Ce/Yb)_N$ varying from 0.3 down to 0.002. However we note that the MREE-LREE segment of the patterns is broadly flat, i.e. the CI-normalized abundances from La to Gd are within error of each other. Within this segment Eu may form negative (or positive) anomalies (with Eu/Eu* ranging from <0.02 to 3). The lowest $(Ce/Yb)_N$ ratios are obtained for the large isolated olivine grains, including the two refractory forsterites analyzed, hereby confirming the results of Pack et al. (2005), with barred olivine chondrules also showing relatively low $(Ce/Yb)_N$ (0.04-0.07), as do some coarse-grained porphyritic chondrules (Fig. 5a). Within the porphyritic chondrules and IOG, a trend of decreasing Ce/Yb with increasing grain size may be seen (Fig. 5b), though scatter is considerable, and GOA do not appear to follow the trend (e.g. relatively coarse-grained GOAs V7 and V45 have $(Ce/Yb)_N$ = 0.2-0.3 while fine-grained GOA V28 has $(Ce/Yb)_N$ = 0.03). It is noteworthy that one analysis in the refractory forsterite V49, located close to the pyroxene-rich margin, gave a significantly shallower slope than the other analyses, with $(Ce/Yb)_N$ = 0.076 (vs. 0.002; with $Lu_N$ = 0.2 vs. 1.2). Although HREE concentrations are well within the range of published values, in contrast our concentration values for LREE ($La_N < 0.15$) are over one order of magnitude lower than previous studies in ordinary chondrites (i.e., $(La)_N > 0.1$ for Alexander (1994) and Ruzicka et al. 2008), but are consistent with analyses of olivine in Allende chondrules reported in an abstract by Kurat et al. (1992). Otherwise, olivine in type I chondrules is similar to the "incompatible-rich" variety of Alexander (1994). The three analyzed type II chondrules lie at the low end of HREE abundances in olivine, as is the case for other refractory elements (Fig. 6).

## *4. Discussion*

### 4. 1 Olivine crystallization

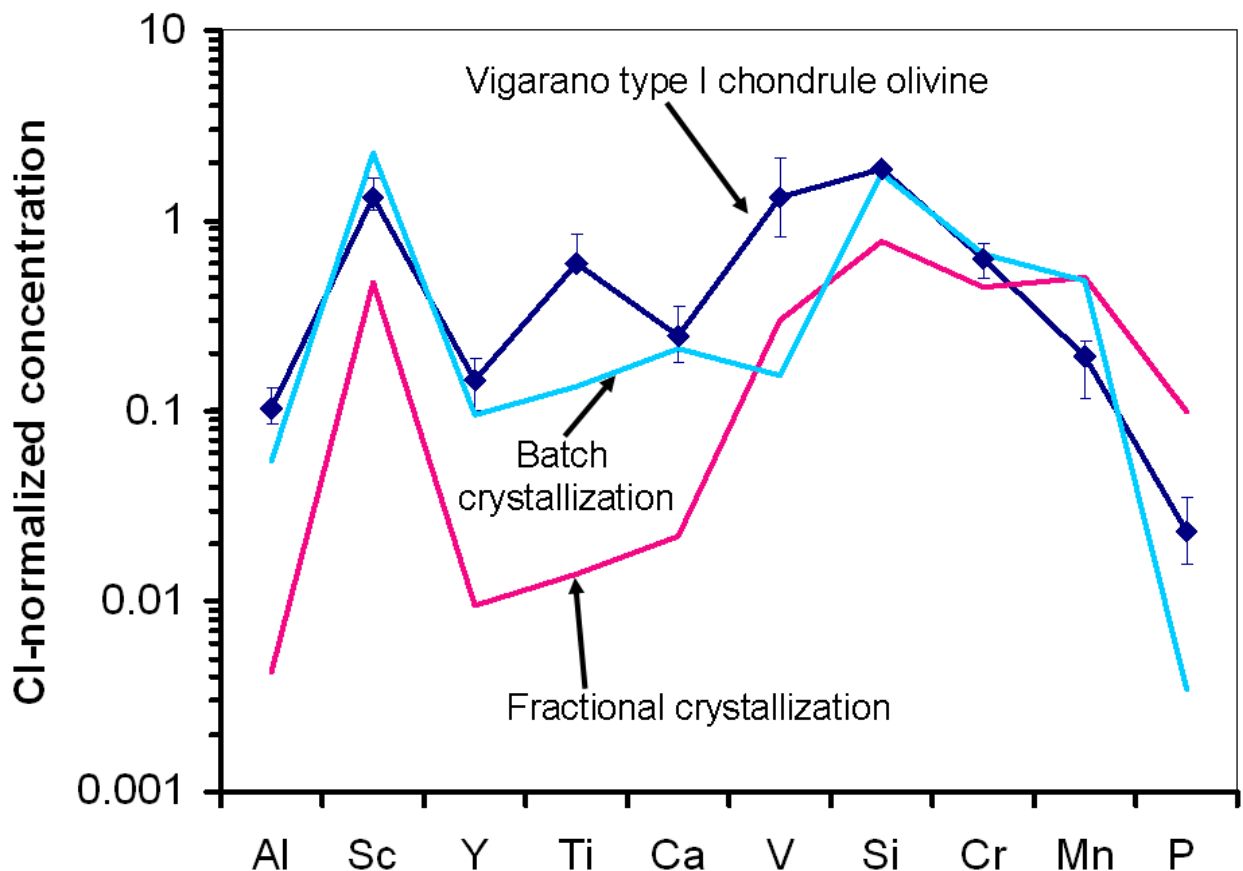

**Figure 7:** Comparison of the predictions for the olivine core composition for fractional and batch crystallization, with the actual composition averaged over the type I chondrules of Vigarano (Renazzo and Acfer 187, being very similar, are omitted for clarity), for the same set of elements as in Fig. 2a. For the fractional crystallization scenario, chondritic abundances are assumed for the bulk composition, while the batch crystallization model uses the measured average composition of the mesostasis. Olivine/melt partition coefficients are drawn from Kennedy et al. (1993), see text for details.

In this section, we investigate the conditions of olivine crystallization as recorded by its trace element budget.

### 4.1.1 Evidence for batch crystallization

In Fig. 7, we compare the average composition of type I chondrule olivine with that calculated from two endmember crystallization models:

(i) a *fractional* crystallization model, where the olivine cores (which is essentially what the LA-ICP-MS analyses access) did not change their composition since they solidified and are thus in equilibrium with the *starting* liquid composition, assumed here to be CI chondritic (Lodders 2003)

(ii) a *batch* crystallization model, where, owing to efficient diffusion, the olivine crystals as a whole maintained equilibrium with the melt now solidified as the chondrule's mesostasis.

For simplicity, we have shown only the results for average compositions in Vigarano, but results for individual chondrules (with analyzed olivine and mesostasis) yield the same conclusions. Olivine/melt partition coefficients were drawn from the experiment "PO49" of Kennedy et al. (1993), conducted at 1525 °C, under 1 atm, with oxygen fugacity 0.5 log unit below the the Iron-Wüstite buffer, using a starting mix of mafic composition, yielding $Fo_{95}$ olivine (as P was not measured, a partition coefficient of 0.1 for that element was taken from Brunet and Chazot (2001)).

From examination of Fig. 7, it is clear that batch crystallization reproduces the observed olivine composition better than fractional crystallization, which yield a composition one order of magnitude too low for refractory elements (although the discrepancy is smaller for the more volatile elements). The difference between the two scenarios is essentially due to the fact that these refractory elements are also incompatible elements, hence the difference between an equilibrium with a starting, chondritic melt, and the late melt represented by the mesostasis. To be sure there are important deviations (variable from chondrule to chondrule (not shown)) for V and Ti (and P, which is less constrained by the literature), but the reducing conditions believed to have prevailed during the formation of type I chondrules (e.g. Zanda et al. 1994), likely rendered them more compatible than in the Kennedy et al. (1994) experiments, as suggested by Ruzicka et al. (2008): this is because $V^{2+}$ and $Ti^{3+}$ have ionic radii (72 pm and 81 pm, respectively) closer to that of the crystallographic sites where they substitute (72 pm, see Wood and Blundy (2003)) than $V^{3+}$ (79 pm) and $Ti^{4+}$ (95 pm), present under more oxidizing conditions. Consistent with this, mesostasis exhibits a clear negative anomaly for V (Fig. 2a) suggesting that it was compatible in olivine.

Conceivably, the conclusion that type I chondrule olivine in the studied meteorites did not form by fractional crystallization could be overturned if (i) starting liquid compositions significantly deviated from chondritic (and in particular were enriched in refractory elements by one order of magnitude, see Fig. 7) or (ii) olivine/melt partition coefficient significantly differed from the chosen data set. As to (i), given that observed bulk chondrule compositions are typically chondritic within a factor of a 2-3 for non-volatile elements (Rubin and Wasson 1987; Misawa and Nakamura 1988; Huang et al. 1996), one would need to assume, similarly to Ruzicka et al. (2008), that chondrule compositions evolved by recondensation during cooling. However, bringing refractory element concentrations >10 x CI down to, say, 2 x CI for a given chondrule would require the chondrule mass to increase by a factor of at least 5 (and this even ignoring the further addition of refractory elements). In other words, most of the chondrule melt would have had to have formed by condensation, contrary to petrographic evidence of chondrule formation largely by melting of precursor material (e.g. Hewins et al. 2005) and lack of significant zoning inside the chondrules (excepting the outer rims). This certainly does not exclude smaller degrees of evaporation/recondensation; in fact, we do observe anticorrelations between refractory and volatile elements (e.g. Ti and Mn in Fig. 3d, similarly to Ruzicka et al. 2008), but as such, they do not enable us to distinguish between vapor fractionation during chondrule formation (e.g. Sears et al. 1996; Hewins et al. 1997) or formation of the precursors. With regards to (ii), while olivine/melt partition coefficients certainly vary depending on physico-chemical condition, it is unlikely that partition coefficients would increase by one order of magnitude for the refractory elements Al, Sc, Y, Ti and Ca while retaining about the same order of magnitude for the more volatile elements V, Si, Cr, Mn, P, judging from the compilation of Bédard (2005) and also the prediction of the Lattice Strain Model (Wood and Blundy 2003), given that no correlation exists between volatility and ionic radius. Therefore, we conclude that olivine in type I chondrules in CR and CV chondrites likely formed by batch crystallization. This is similar to the conclusion drawn by Huang et al. (1996) for type I chondrules in ordinary chondrites on the basis of their observed crystallization sequence, and this is certainly consistent with the lack of minor element zoning in type I chondrules in CR and CV chondrites (Brearley and Jones 1998; neither our EMP or LA-ICP-MS data reveal zoning at crystal or chondrule scales).

Our conclusion that most type I chondrules we analyzed underwent batch crystallization implies that the thermal event was sufficiently long to allow homogenization of olivine crystals by cation diffusion. An estimate for the cooling rate corresponding to a diffusion length $a$ is given by (Dodson 1973):

$$\frac{dT}{dt} = -2\frac{D(T)}{a^2}\frac{RT^2}{E_a}, \quad (1)$$

with $D(T)$ the diffusion coefficient at temperature $T$ (defined as that when crystal growth occurs), $E_a$ the activation energy (in J/mol) of the diffusion process and $R = 8.314$ J/mol the gas constant. For a batch crystallization scenario to hold, $a$ must be at least of order the crystal radius, so taking $a = 50$ μm should yield an upper limit to the cooling rate. According to the review by Chakraborty (2010) who compiled diffusion data for olivine for an extensive array of elements (Be, V, Ti, Na, Zr, Ca, Cr, Si, O, Mn, Ni, Co, REE, H, Li), $E_a$ is typically 300 kJ/mol (within a factor of two) and diffusion coefficients for temperatures of 1500-2000 K (encompassing the range of chondrule peak temperatures reported by Hewins et al. 2005) seem to be at any rate smaller than about $10^{-14}$ m²/s for all elements they reviewed (except H and Li). This implies an upper bound of order ~10 K/h.

While we favor batch crystallization for type I chondrules in CV and CR chondrites, our limited data set for more FeO-rich chondrules is still consistent with fractional crystallization for them, as suggested e.g. by their minor element zoning (e.g. Jones 1990; Huang et al. 1996): indeed, olivine from the three analyzed type II chondrules tends to have low incompatible refractory element abundances (Fig. 6). However, less mafic compositions of the melt should *increase*, in general, olivine/melt partition coefficients (Bédard 2001) and hence the abundance of these elements in olivine were batch crystallization to hold equally for them. This implies that olivine cores are likely in equilibrium with the bulk chondrule composition, i.e., the initial melt composition, rather than the final, incompatible element-enriched one, presumably due to fast cooling which prevented homogenization of olivine crystals (100-1000 K/h; Jones 1990). More data on type II chondrules are certainly needed.

## 4.1.2 Partitioning of the most incompatible elements

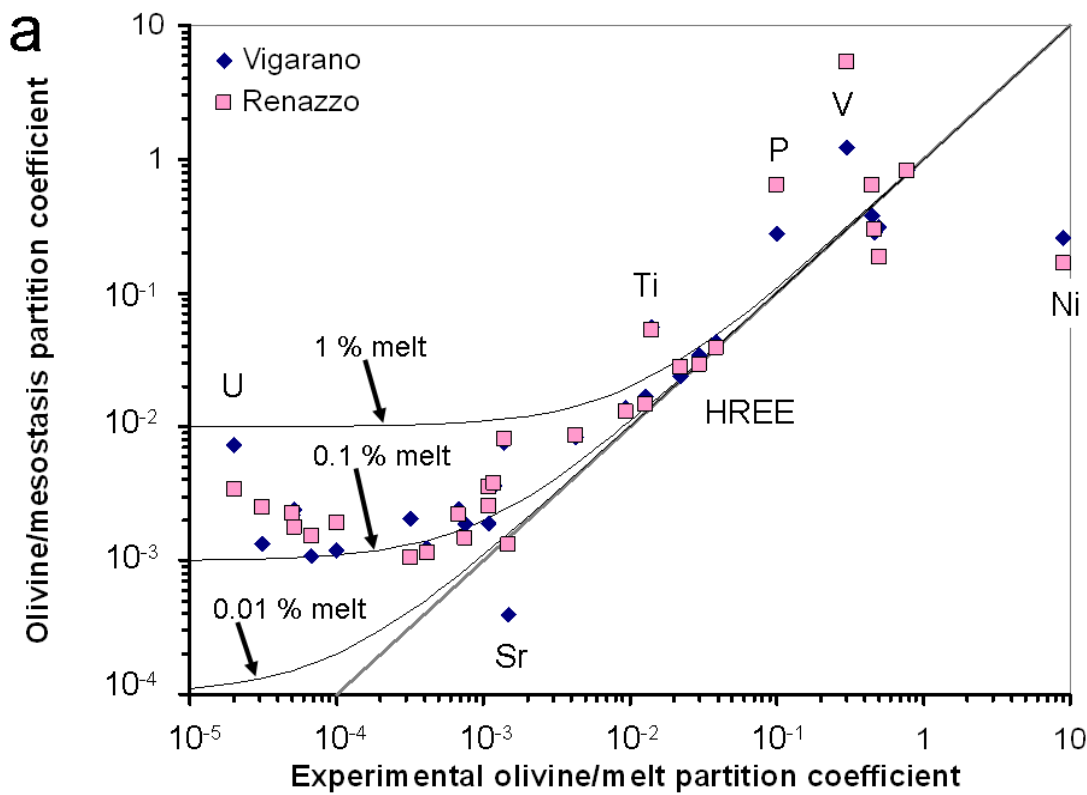

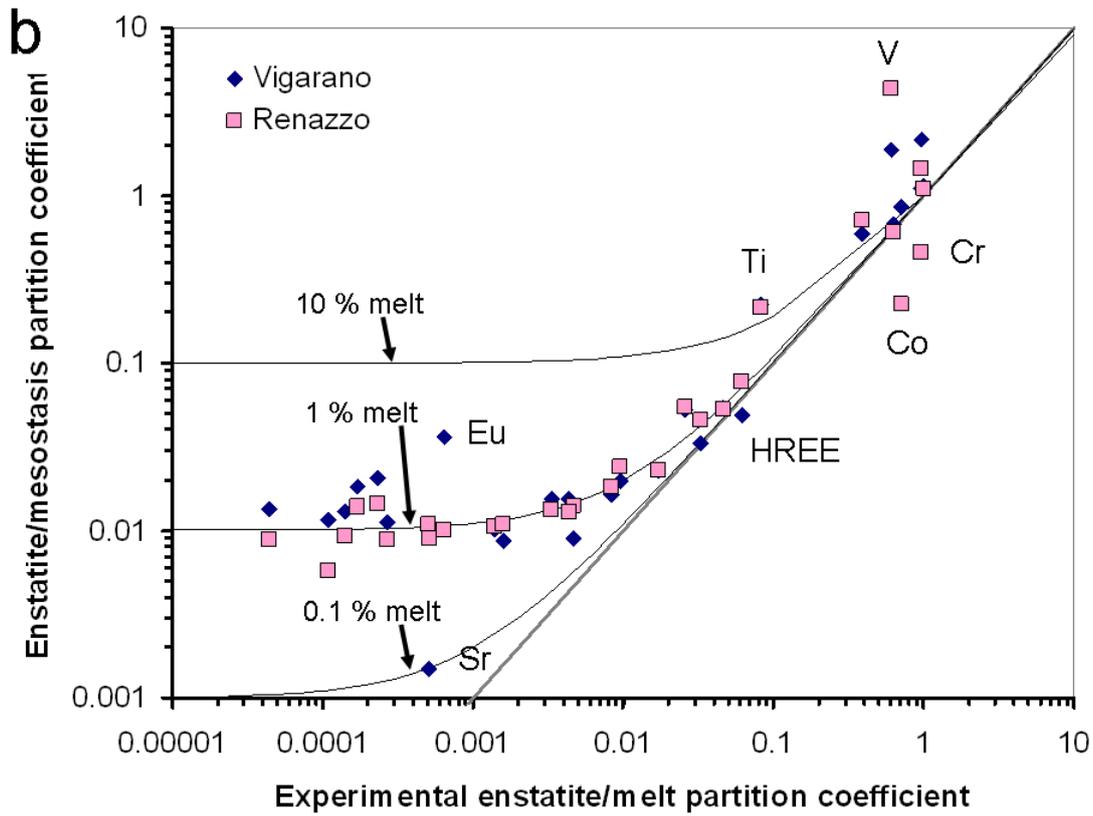

**Figure 8:** (a) Comparison of experimental olivine/melt partition coefficients (from Kennedy et al. 1993) with observed olivine/mesostasis partition coefficient (calculated here from the average compositions of those phases in type I chondrules in Vigarano and Renazzo). The heavy line corresponds to a 1:1 agreement, and thin lines are calculations assuming different proportions of melt contamination in the analyzed olivine (using equation 2). (b) Same for enstatite, with the experimental enstatite/melt partition coefficients drawn from the run RPII 45 of Kennedy et al. (1993).

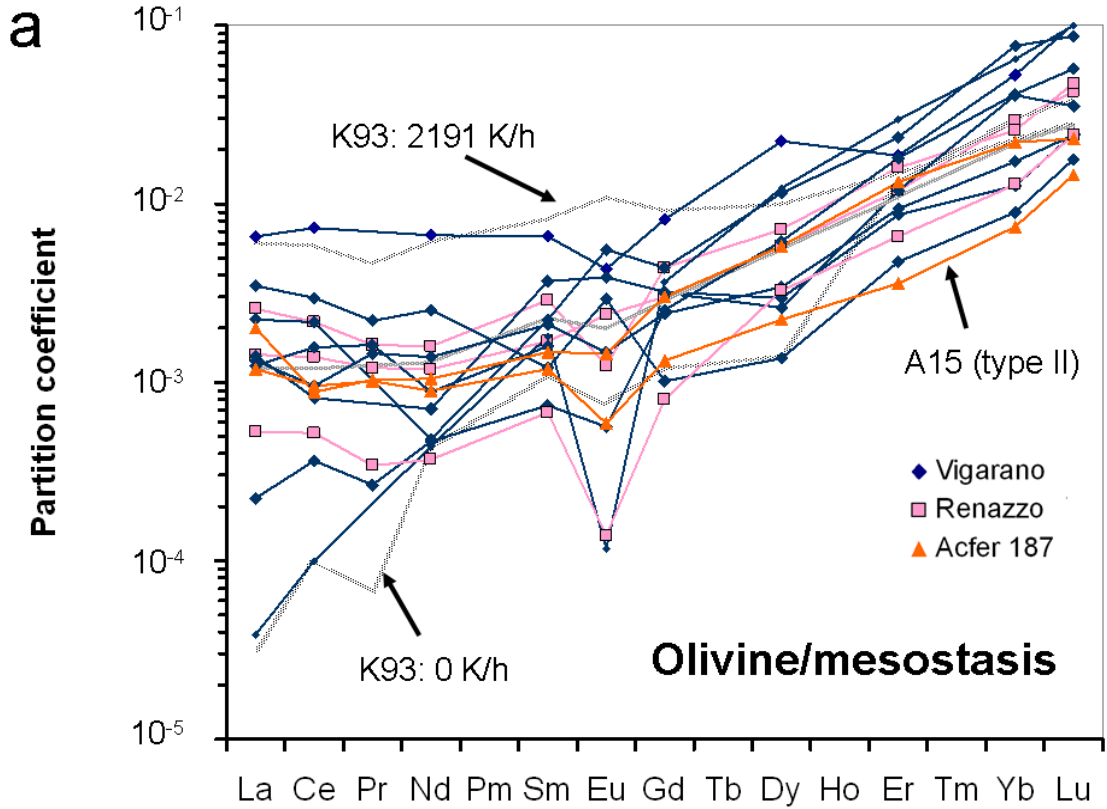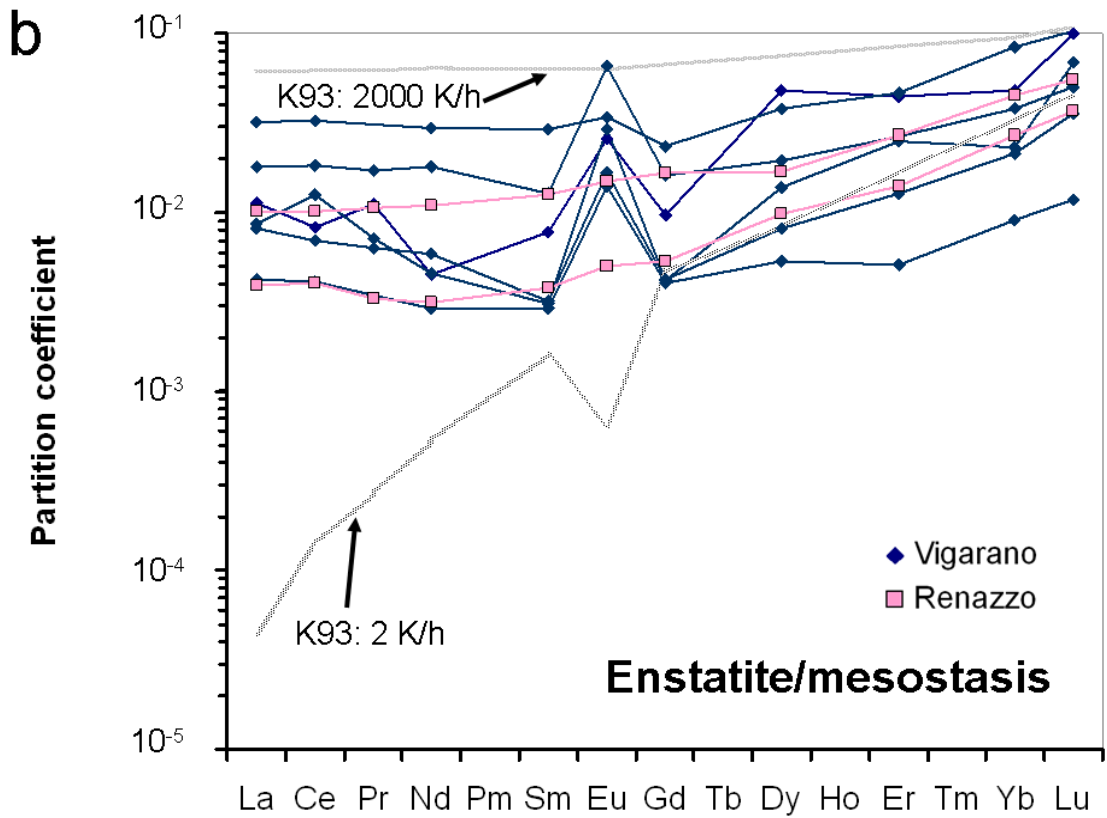

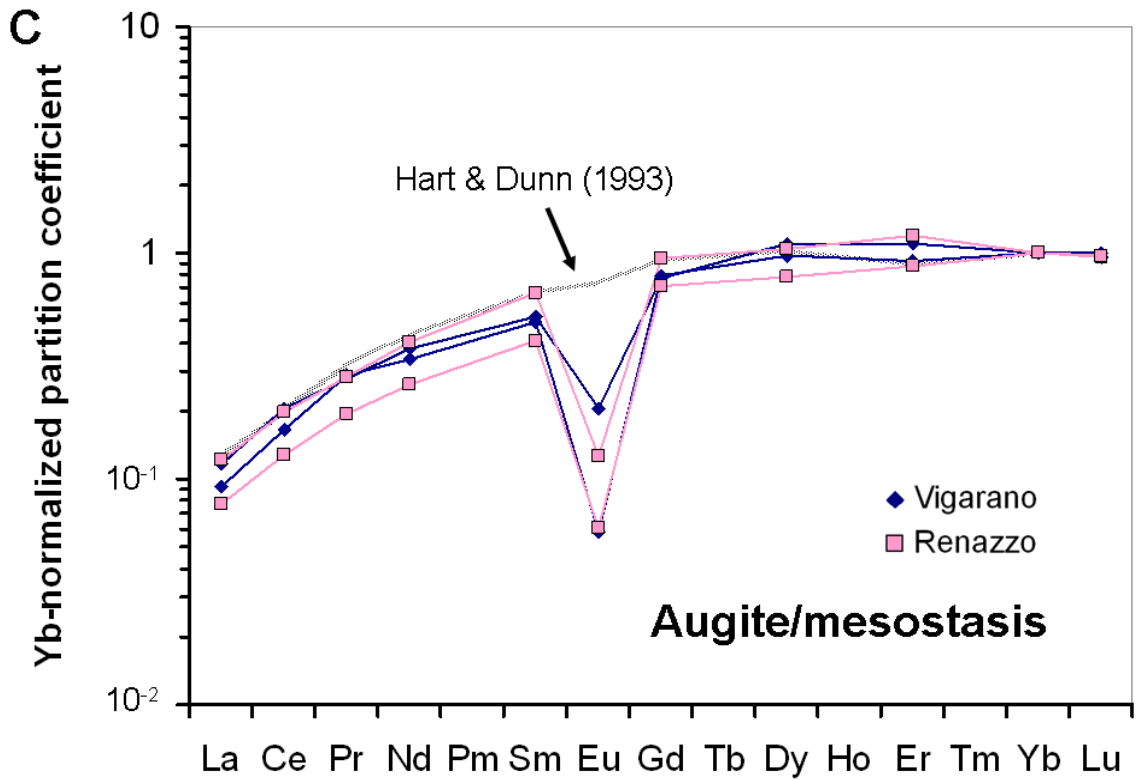

**Figure 9:** Partition coefficients of (a) olivine, (b) low-Ca pyroxene and (c) Ca-rich pyroxene with mesostasis. For Ca-rich pyroxene, partition coefficients are normalized to that of Yb to cancel dependence (to first order) on Ca content (see Fig. 6b). For each chondrule, olivine, low-Ca pyroxene and mesostasis compositions are means over the analyses performed on the chondrule. In (a) and (b), we have plotted in gray experimental values reported by Kennedy et al. (1993) ("K93")(a) for their PO49 ("K93: 0 K/h") and BO30 ("K93: 2191 K/h") experiments and (b) for their RPII 45 ("K93: 2 K/h") and RP36 ("K93: 2000 K/h") experiments. In (c), we have plotted in gray clinopyroxene/melt partition coefficients reported by Hart and Dunn (1993). In (a), the single datum from a type II chondrule (A15), is highlighted.

In Fig. 8a, we compare measured olivine/mesostasis partition coefficients with experimentally determined olivine/melt partition coefficients for all elements (with the aforementioned data set from Kennedy et al. 1993), and Fig. 9a presents olivine/mesostasis partition coefficients for individual chondrules. Consistent with our above conclusion that olivine formed by batch crystallization in type I chondrules, for elements with experimental partition coefficients larger than $10^{-3}$-$10^{-2}$ (including the HREE), the two partition coefficients generally agree, within a factor of a few at most, except for the already mentioned outliers Ti, V, P, as well as Ni which may also be subject to redox effects (Ruzicka et al. 2008). For nominally more incompatible elements, however, observed olivine/mesostasis partition coefficients appear to converge toward about $10^{-3}$, which manifests itself in particular in the flattening of the REE patterns for LREE in Fig. 9a. A possibility could be that analyzed olivine incorporates some fraction of a melt-like component (e.g. Kennedy et al. 1993; Hiraga and Kohlstedt 2009). In that case, assuming that the composition of this purported component is that of the liquid in equilibrium with the olivine,, the apparent (olivine + inclusions)/melt partition coefficient would be given by:

$$D_{apparent} = D_{real}(1-\varphi) + \varphi, \quad (2)$$

with $D_{real}$ the real (equilibrium) silicate/melt partition coefficient (not to be confused with the diffusion coefficients discussed in the previous subsection) and $\varphi$ the mass fraction of the incorporated melt component. While for $D_{real} > \varphi$, $D_{apparent}$ hardly differs from $D_{real}$, for more incompatible elements, $D_{apparent} \approx \varphi$, as illustrated in Fig. 8a. From examination of Fig. 8a, this would imply a typical (but variable, see Fig. 9a) melt fraction of $\sim 10^{-3}$ in olivine. As SEM observations and signal filtering exclude the presence of >1 µm inclusions in the undiscarded analyses, the hypothetical trapped melt would need to exist as smaller pockets and TEM studies would be worthwhile to test this idea: If, for example, the hypothetical melt inclusions are assumed to be 100 nm in diameter, their required number density in the olivine section would be $\sim 0.1$ µm$^{-2}$. It is notable, however, that the olivine/mesostasis partition coefficient tends to increase slightly for the lightest REE, hence a "spoon shape" in the patterns of Fig. 9a, instead of remaining constant as expected if the incorporated component were strictly mesostasis-like in composition. Alternatively, the enhanced partition coefficient of very incompatible elements could be a crystal-lattice-scale kinetically controlled disequilibrium effect due to rapid crystal growth (e.g. Albarède 2002; Watson 1996). Yet another possibility, suggested by Kurat et al. (1992), could be that the olivine originally had a flat REE pattern, e.g. inherited from condensation, and was only partially equilibrated with the melt.

Regardless of how very incompatible elements are actually incorporated in the analyzed olivine (viz. as submicrometric inclusions or in the crystal lattice), which factors control the (variable) extent of this incorporation? To tackle this question, we return to the variations of the Ce/Yb (a proxy for this effect, as Ce is a very incompatible LREE) with petrography as illustrated in Fig. 5, and in particular, the apparent anticorrelation between Ce/Yb and olivine grain size (Fig. 5b), with the coarsest IOG exhibiting (Ce/Yb)$_N$ values close to the equilibrium partitioning prediction (0.003 according to Kennedy et al. 1993).

A concern could be that this trend may be an analytical artefact, with the cleanest LA-ICP-MS analyses coming from the largest crystals, which would also be least affected by contaminations of various natures. The trend is however unlikely to result from diminishing detection limits with increasing spot size: Indeed, the 99 % confidence detection limit for Ce was typically 1 ppb for 26 µm spots down to 0.1 ppb for 102 µm spots, to be compared with the 0.7-70 ppb range of measured concentrations. Then, were small and large crystals actually intrinsically similar, the use of larger LA-ICP-MS spots for the latter would be expected to *average* concentrations from smaller scales (assuming some heterogeneity of olivine crystals), yielding a smaller range within that of smaller olivine analyses, contrary to the overall decrease observed in Fig. 5. Thus, while analytical artefacts cannot be excluded, we are led to conclude that the trend in Fig. 5, however scattered, is real. How can it be then interpreted?

We first note that this trend runs counter to the Kurat et al. (1992) hypothesis of partial equilibration of olivine with a flat CI-normalized trace element pattern. Indeed the coarsest olivine crystals would be expected to be least equilibrated because the larger diffusion length to be attained. Thus, even for the most incompatible elements, we believe that olivine/mesostasis partition coefficients retain *per se* no memory of the chondrule precursor history and must be understood in terms of igneous (or subsolidus) processes during chondrule formation.

In chondrules, crystal size is determined by cooling rate and also—and primarily—by the number of nuclei surviving the heating event: For a given peak temperature, the longer the heating, the coarser the grain size (Lofgren 1996; Hewins and Fox 2004). So crystal size may be considered as a proxy for the overall duration of the chondrule-forming event. In addition, barred olivine chondrules, which were not included in Fig. 5b because of the ambiguity in defining a grain size for them, likely represent heating times intermediate between porphyritic chondrules and IOG, consistent with the trend seen in Fig. 5a, as likely all nuclei (or all but one) were destroyed in these chondrules during heating. This suggests the following interpretation of the anticorrelation between crystal size and Ce/Yb: The finer-grained a chondrule is, the shorter the thermal event it experienced, and thus the more trace element partitioning deviated from equilibrium. A cooling-rate sensitivity of olivine/melt partition coefficients was observed in the experiments of Kennedy et al. (1993), where partition coefficients for LREE increased to about $10^{-2}$ for LREE (compared to $3.1 \times 10^{-5}$ for La at equilibrium) in a rapid cooling (2193 K/h) run.

Another possibility would be to invoke post-crystallization redistribution of trace elements, e.g. in a second heating events if olivine crystals are "relicts" from an earlier generation, which would affect small crystals most easily (as the diffusion length is shortest). An illustration for this could be the olivine analysis at the edge of refractory forsterite V49 (Fig. 1a), which, as we mentioned in the "Results" section, yielded less REE fractionation than analyses in the interior. Conceivably, this IOG could have been remelted at the periphery (perhaps in the event that formed its thin enstatite rind, see Section 4.2), whose trace element budget would have been reset. Similar redistributions could explain why GOAs, which have been interpreted as relicts by Libourel and Krot (2007), do not appear to follow the trend, and more generally explain part of the scatter in Fig. 5b.

If the incorporation of very incompatible elements is kinetically controlled, the LREE/HREE fractionation could in principle help estimate the cooling rate of chondrules. Using this approach, the experiments of Kennedy et al. (1993), plotted in Fig. 9, would suggest cooling rates of the order of 1000 K/h for most chondrules (be they of type I or II), similarly to Alexander (1994) and

Ruzicka et al. (2008). This would stand in stark contrast to the upper limit of about 10 K/h we have inferred above from evidence of batch crystallization—unless trace element diffusivities in olivine were somehow much higher than all the data reviewed by Chakraborty (2010). However, in the uncertainty regarding the microscopic process(es) controlling their incorporation and the degeneracy possibly introduced by post-crystallization resetting, as we have just discussed, a calibration of cooling rates based on the Kennedy et al. (1993) data may not be robust. For instance, Alexander et al. (1998) set to reproduce the Kennedy et al. (1993) experiments but found a considerably weaker sensitivity to cooling rates. On the other hand, weakly fractionated REE patterns ($0.1 \leq (Ce/Yb)_N \leq 1$) in olivine are observed in terrestrial peridotites (e.g. Hiraga and Kohlstedt 2009), possibly the acapulcoite Monument Draw (McCoy et al. 1996), and SNC meteorites such as nakhlites and Chassigny (Wadhwa and Crozaz 1995), which are slowly cooled rocks (< 0.1 K/h for Nakhla and Governador Valadares, for example, cf Mikouchi and Miyamoto 2002). Thus, the variation of apparent partition coefficients under disequilibrium conditions, although seen in our data, is not understood clearly enough to base a calibration of cooling rates, and we are thus led to prefer the low cooling rates estimates from diffusion data.

### 4.1.3 Implications for the origin of olivine

In this subsection, we discuss some implications of the above considerations on the genesis of olivine in type I chondrules in CV and CR chondrites.

The first, straightforward implication is that olivine in the analyzed chondrules has an igneous origin, as evidenced by reasonable agreement between olivine/mesostasis partition coefficients and equilibrium olivine/melt value. Equilibrium partitioning is in particular satisfied even for the most incompatible elements (e.g. LREE) for large IOG, including refractory forsterites, confirming the igneous origin inferred by Pack et al. (2005) for the latter, at variance with Weinbruch et al. (2000) who had favored condensation. In no way does this of course exclude a non-igneous (e.g. condensation) origin for chondrule precursors.

We have inferred above an upper limit of about 10 K/h for the cooling rate recorded by olivine in type I chondrules. Cooling rates around 10 K/h have been found to be consistent with PO chondrule textures (Wick et al. 2010) and, as regards CR chondrites specifically, Humayun (2012) estimated cooling rates of 0.5-50 K/h (around $1473 \pm 100$ K) from the diffusion of Cu and Ga in coarse metal grains in Acfer 097 (CR2). Similar cooling rates have already been inferred for granular chondrules from Allende from the wavelengths of pigeonite-diopside intergrowths and the microstructure of calcic plagioclase, both established at temperatures around 1500 K (Weinbruch and Müller 1995). They are also comparable to those reproducing the texture of igneous CAIs, in the range 0.5-50 K/h (Stolper and Paque 1986; Apai et al. 2010), and reminiscent of the timescales (days) required to produce GOA from fine-grained olivine (Whattam et al. 2008). All this suggests that the solids thermally processed in the accretion disk exhibited a somewhat continuous spectrum of cooling rates, extending from 1 K/h to the 1000 K/h inferred for type II chondrules (Hewins et al. 2005). This leaves open the possibility (raised e.g. by Jones et al. 2000) that chondrule- and igneous CAI-melting events arose from the same physical process (acting on different precursors), provided it was variable enough (at least in terms of the ensuing cooling rates) over the course of the accretion disk lifetime. We note that cooling rates spanning the whole range 10-1000 K/h were found in one-dimensional shock wave models by Morris and Desch (2010).

Libourel and Krot (2007) suggested that precursors of PO chondrules, best approximated by GOAs, were mantle debris of an early generation of planetesimals, which would account for the numerous triple junctions of the interlocking olivine crystals. Would our trace element data be consistent with such an hypothesis? While evidence for batch crystallization would certainly be consistent with a long residence in a planetary mantle, provided part of the chondrule mesostasis was cogenetic with the olivine, we have seen that the REE patterns of olivine differs from equilibrium fractionation prediction, except for the coarsest crystals. However, olivine in most differentiated meteorites display highly fractionated REE consistent with equilibrium partitioning: This includes pallasites (Minowa and Ebihara 2002), ureilites (Guan and Crozaz 2000), brachinites (Wadhwa et al. 1998), angrites (Floss et al. 2003) lherzolitic shergottites (Usui et al. 2010 but see Lin et al. 2005) and the lunar olivine gabbro NWA 2977 (Zhang et al. 2010). Conceivably, olivine in type I chondrules may have originally had steep REE patterns, but would have had its trace element budget reset subsequently, as we have mentioned, but we find this unlikely given that only one analyzed GOA (V28) out of 7 has a $(Ce/Yb)_N$ ratio below 0.1 (0.04). Nonetheless, the fact that olivine in terrestrial peridotites and cumulate SNC displays more limited HREE/LREE fractionation (Hiraga and Kohlstedt 2009; Lin et al. 2006; Wadhwa and Crozaz 1995) is evidence that conditions exist under which planetary olivine may have REE signature different from equilibrium. We finally note that our petrographic observations, such as the continuity of GOA with fine-grained and/or close-packing textures, or existence of "non-relict" GOA such as chondrule V4 (Fig. 1), would appear to favor a nebular origin for granoblastic textures, although experiments by Whattam et al. (2008) were not entirely successful at reproducing them. Overall, our data provide no evidence for a planetary origin of olivine in type I chondrules, but do not rule it out.

## 4.2 Pyroxene crystallization

We now turn our attention to pyroxene. As can be seen in Fig. 8b and 9b, a departure of observed enstatite/mesostasis partition coefficients from experimental enstatite/melt partition coefficients can be seen, similarly as for olivine, for the most incompatible elements. These typically have observed enstatite/mesostasis partition coefficients of the order $10^{-2}$, which is one order of magnitude greater than for olivine. If interpreted to result from incorporation of a melt-like component, then a level of contamination of 1 %, at the submicron scale, would be required (perhaps in the shrinkage cracks that pervade enstatite; see Brearley and Jones (1998)). The positive europium anomaly often exhibited by enstatite, correlated with La/Sm (Fig. 4c), could point to a more plagioclase-like character of this purported component. As to augite, no enhanced incorporation of very incompatible elements is observed (Fig. 9c), likely because the expected partition coefficients are higher than for enstatite (Hart and Dunn 1993), but the negative europium anomalies may indicate that part of Eu is in the divalent state (e.g. Jones and Layne 1997) and preferentially incorporated in the feldspar crystallites of the mesostasis.

Not only are LREE more abundant in enstatite than in olivine, they also never approach equilibrium partitioning with mesostasis (Kennedy et al. 1993), unlike the case of olivine (see Section 4.1.2). This suggests that enstatite records rapid cooling rates. A more quantitative estimate was drawn by Jones and Scott (1989) from the crystallographic structure of enstatite, namely the fact that enstatite is in monoclinic, rather than orthorhombic form: a lower limit of 100-1000 K/h at the temperature of inversion from protoenstatite (the high-temperature form of enstatite), about 1000 °C, was inferred. This contrasts with the upper limit of about 10 K/h we have derived above for olivine crystallization, at higher temperatures (with peaks at 1400-1750 °C; Hewins et al. 2005). The only way we can think of to avoid this difference would be an enhancement of trace element diffusivities in olivine relative to all the data of the Chakraborty (2010) review. If this difference is real, this could imply that the cooling history of chondrules was highly nonlinear, with a stage of "quenching" after olivine crystallization (e.g. Weinbruch and Müller 1995), which differs e.g. from the shock wave model temperature curves, where cooling tends to slow down after peak heating (e.g. Morris and Desch 2010). Alternatively, the pyroxene formation event could have been distinct from that of olivine crystallization (i.e. olivine would be relict; e.g. Libourel et al. 2006).

As we saw earlier, there appears to be an anticorrelation between REE concentration in enstatite and the overall mode of this silicate in chondrules (Fig. 4a), despite scatter in part attributable to the uncertainty of equating 2D mode with 3D mode (e.g. Hezel and Kiesswetter 2010). In fact, this trend is tighter when REE abundances are expressed as enstatite/olivine partition coefficients (Fig. 10), although no trend would be expected if olivine and enstatite were fully equilibrated, given that partitioning behaviors of trace elements are quite similar in both silicates in terms of Lattice Strain

Model parameterization (Wood and Blundy 2003). We propose that this trend could be explained in the framework of the Libourel et al. (2006) model (see also Hezel et al. 2003), where enstatite (mainly near the chondrule's edge) forms, partly at the expense of olivine, as a result of the addition of silica to the chondrule melt (which Libourel et al. (2006) ascribed to gas-melt interaction). Indeed silica enrichment in the melt would induce, to first order, a dilution of all elements other than Si and O in it, e.g. Ca and Al as witnessed by Libourel et al. (2006), but also trace elements such as the REE. Thus, with increasing progress of silica addition (to which the enstatite mode would be a proxy) — be it by gas-melt interaction or another process —, pyroxene in equilibrium with the melt should have lower concentration of these trace elements. The trend would be exacerbated by the negative effect of falling Ca concentrations (because of the dilution) on enstatite/melt partition coefficients (see Wood and Blundy 2003) and indeed the trend for enstatite (Fig. 4a) is steeper than for mesostasis (Fig. 4b).

The emerging picture here is thus that of a rapid pyroxene-forming event, presumably promoted by silica addition, subsequent to olivine crystallization (be the pyroxene forming event distinct or in continuation to that of olivine crystallization). Olivine would have in general preserved its prior chemistry, although some resetting may have occurred (see Section 4.1.2). Experimental work reported by Soulié et al. (2012) moreover suggests that olivine dissolution in a melt compositionally similar to the present-day mesostasis could proceed at rates faster than 100 µm/h, implying a rapid cooling during pyroxene formation to account for the preservation of olivine phenocrysts; also, the fact that enstatite is restricted to the margin would limit the diffusion timescale of Si in the melt: if a Si diffusivity of $10^{-11}$ m²/s (Zhang et al. 2010) is assumed, diffusion over 100 µm is effected within about 1 h, for instance.

We note that in such a scenario, the composition of mesostasis would have evolved out of pace with olivine, so that strictly speaking, the olivine/mesostasis equilibrium we have inferred in Section 4.1.1 would not have been sustained. However, given the limited variation of mesostasis REE contents (which only show a shallow trend as a function of pyroxene mode in Fig. 4b), the effect on olivine/mesostasis partition coefficients would have been minimal, especially with regards to the uncertainties of their theoretical values, and would hardly affect our argument in favor of batch crystallization, as opposed to fractional crystallization.

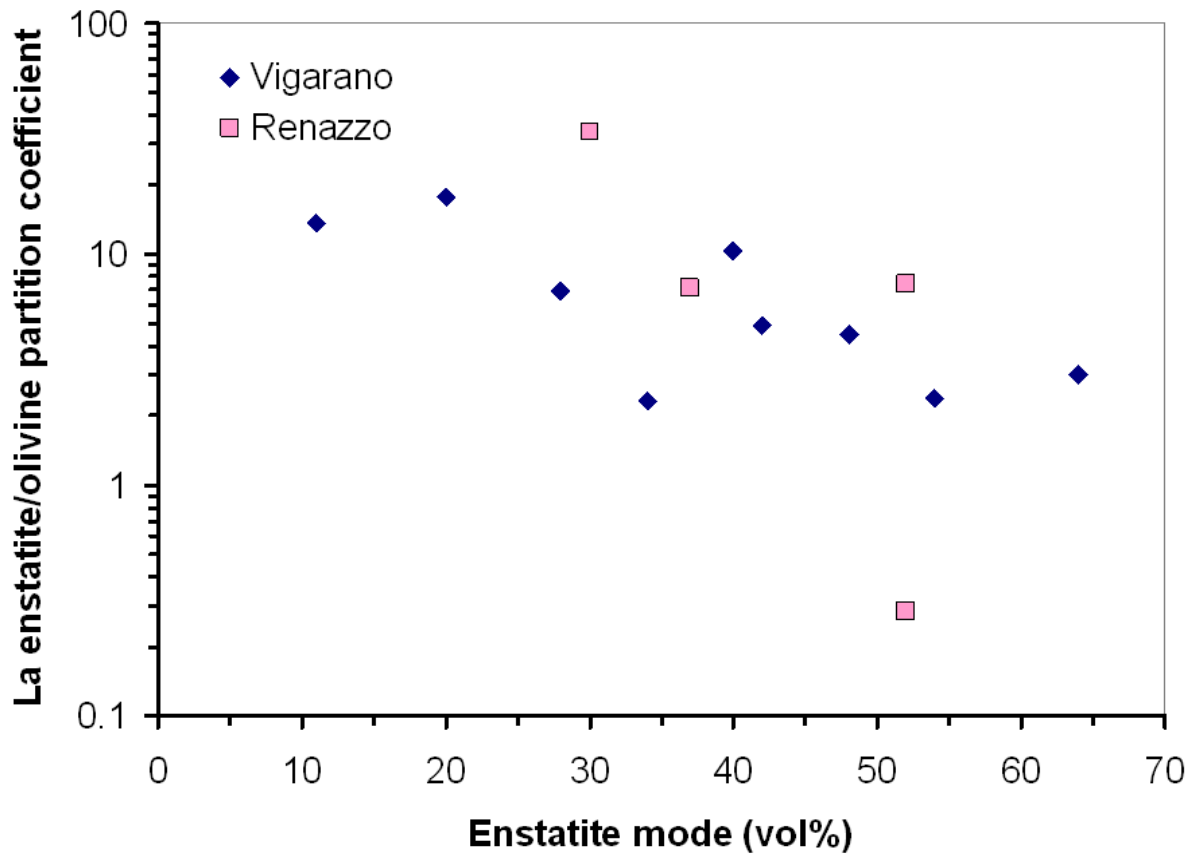

**Figure 10:** Plot of the olivine/low-Ca pyroxene partition coefficient for La versus low-Ca pyroxene mode.

## *5. Summary and outlook*

We have carried out LA-ICP-MS trace element analyses (with special emphasis on REE) of olivine, pyroxene and mesostasis in chondrules of various textures of Vigarano (CV3), Renazzo and Acfer 187 (CR2). Results are broadly similar in all three carbonaceous chondrites and may be summarized as follows:

1° Olivine and pyroxene have fractionated REE, with HREE concentrations at 0.1-1 x CI for both silicates and La concentrations at 0.0008-0.15 x CI and 0.05-0.5 x CI for olivine and pyroxene, respectively.
2° Mesostasis has flat REE patterns, mostly at 10-20 x CI.
3° Small olivine grains tend to have higher LREE/HREE ratios than coarser grains.
4° Low-Ca pyroxene in the most pyroxene-rich chondrules tend to have the lowest REE contents.

From these data, the following interpretations were drawn:

1° The fact that, for type I chondrules, olivine/mesostasis partition coefficients for most elements are consistent with experimental data indicate efficient diffusion of these elements through olivine crystals (counter to a fractional crystallization scenario). From literature values of diffusion coefficients, this seems to require cooling rates no higher than 1-100 K/h. Thus, nebular solids seem to have recorded a more or less continuous array of cooling rates, from those experienced by igneous CAIs to those inferred for type II chondrules (up to 1000 K/h).

2° The correlation between LREE/HREE fractionation and crystal size for olivine indicates that olivine/melt REE partitioning is controlled by kinetic effects and/or the incorporation of melt-like inclusions in olivine, whether during initial growth of the crystals or subsequent reheating. An igneous origin for refractory forsterites is confirmed by their olivine/mesostasis REE partition coefficients consistent with equilibrium igneous partitioning. As most olivine crystals deviate from equilibrium LREE/HREE fractionation, we do not favor a planetary origin for them.

3° The anticorrelation between REE content in low-Ca pyroxene and pyroxene mode in chondrules may be understood as arising from a dilution following addition of silica (promoting pyroxene crystallization) in the melt, as in the Libourel et al. (2006) model. The high cooling rates necessitated by low-Ca pyroxene suggest either a nonlinear cooling history, or that the formation of the pyroxene-rich margins was an event distinct from that which formed the enclosed olivine.

## *Acknowledgments*

We thank the Associate Editor Christine Floss, Dominik Hezel, and an anonymous reviewer for their reviews that greatly improved the logical clarity of the manuscript and in particular the discussion section. We are grateful to the Programme National de Planétologie and the Institut Universitaire de France for their financial supports, NASA and the Museum National d'Histoire Naturelle for the loan of the samples.